\documentclass[preprint,review,12pt]{elsarticle}
\journal{XXX}

\usepackage{amssymb}
\usepackage{amsmath,bm}
\usepackage{afterpage}
\usepackage{subcaption}
\usepackage{colortbl,booktabs}
\usepackage{tabularx}
\usepackage{threeparttable}
\usepackage{multicol,multirow}
\usepackage [font=footnotesize,labelfont=bf]{caption}
\usepackage{rotating}
\usepackage{tikz}
\usepackage{hyperref}
\usepackage[section]{placeins}
\usepackage{afterpage}

\usepackage{lineno}

\usepackage{color}
\definecolor{R}{rgb}{1, 0, 0}
\definecolor{G}{rgb}{0, 0.5, 0}
\definecolor{B}{rgb}{0, 0, 1}

\journal{Aerospace Science and Technology}
\makeatletter
\def\ps@pprintTitle{%
  \let\@oddhead\@empty
  \let\@evenhead\@empty
  \def\@oddfoot{} 
  \def\@evenfoot{} 
}
\makeatother

\begin{document}

\begin{frontmatter}

\title{{Denoising diffusion and latent diffusion models for physics field simulations}}
\author[address1]{Yuan Jia }

\author[address1]{Chi Zhang}

\author[address2]{Hao Ma \corref{mycorrespondingauthor}}
\cortext[mycorrespondingauthor]{Corresponding author.}
\ead{mahao@zua.edu.cn}
\author[address1]{Qiao Zhang}
\author[address1]{Kai Liu}

\author[address1]{Chih-Yung Wen \corref{mycorrespondingauthor}}

\ead{chihyung.wen@polyu.edu.hk}

\address[address1]{Department of Aeronautical and Aviation Engineering, \\The Hong Kong Polytechnic University, China}
\address[address2]{School of Aerospace Engineering, Zhengzhou University of Aeronautics,\\ Zhengzhou, 450046, China}

\begin{abstract}
Accurate prediction of physical fields is critical in various engineering applications, including thermal management in electronic systems, airfoil shape optimization in aerospace, and flow field control in hypersonic vehicles. This study employs the  Denoising Diffusion Probabilistic Models (DDPMs) for predicting the temperature field caused by the thermal diffusion, and the flow fields spanning from incompressible to hypersonic regimes.  A conditional DDPM framework is first validated with a steady-state thermal diffusion problem by predicting the temperature distribution around a plate with holes. Strong agreement with ground truth data is shown with an average error of approximately 0.013 for plates with a central circular hole. The model also delivers high accuracy in critical regions, such as near the inner circular or square holes. Its performance is further evaluated on incompressible flow around an airfoil and hypersonic flow over a compression ramp, confirming robust predictive capability across diverse flow conditions. Additionally, a latent-space implementation of DDPM is introduced, which employs an Autoencoder (AE) for dimensionality reduction and reconstruction of the physical data. The resulting Latent Diffusion Model (LDM) maintains reconstruction quality comparable to the standard DDPM while substantially reducing the computational cost of the diffusion training process. When applied to hypersonic flow over a compression ramp in the original parameter space, LDM predictions align well with ground truth, achieving a deviation of only 4.28\% in separation length estimation. This work confirms the high predictive accuracy of the DDPM framework and highlights the efficiency gains from performing diffusion in a learned latent space. The findings establish an efficient framework for high fidelity generative modeling of complex thermal/flow fields.

\end{abstract}


\end{frontmatter}

%
%
\section{Introduction}
\label{sec1}
Computational Fluid Dynamics (CFD) serves as a fundamental tool in modern scientific and engineering fields \cite{blazek2015computational,udoewa2012computational,bhatti2020recent,zakaria2018computational,jia2021power}, with extensive applications in aerospace \cite{spalart2016role,mani2023perspective}, automotive design \cite{dhaubhadel1996cfd}, energy systems \cite{bhatti2020recent}, and environmental science \cite{zhang2024accelerating}. High fidelity simulation of physical fluid fields is essential for analyzing complex physical phenomena and guiding engineering design. However, obtaining accurate and high-resolution flow fields remains challenging in scenarios involving complex geometries, multiphysics coupling, and broad flow velocity regimes—from incompressible to hypersonic flows \cite{longo2007challenge,keyes2013multiphysics}.  Although traditional CFD methods such as the Finite Volume Method (FVM) and Finite Difference Method (FDM) can achieve high accuracy, they require substantial computational costs, which limit their use in real-time control and optimization processes \cite{jeong2014comparison}. For instance, a typical optimization loop may demand thousands of CFD evaluations, each consuming several hours of computation, resulting in prohibitively long total computation times.

To address these limitations, deep learning methods have gained popularity and have been successfully applied across various engineering domains. Notable applications include indoor airflow prediction \cite{zhang2024enhancing}, temperature distribution estimation in electronic devices, airfoil flow field reconstruction \cite{chen2023towards}, and hypersonic flow modeling  \cite{jia2025global,jia2025prediction}, significantly advancing the integration of machine learning in fluid dynamics. Recently, with the rise of Generative Artificial Intelligence (GenAI), Denoising Diffusion Probabilistic Models (DDPMs) \cite{song2020denoising,li2024synthetic} have attracted increasing attention due to their ability to reconstruct high-quality flow fields from limited training data. Compared to other generative models like Generative Adversarial Networks (GANs), DDPMs offer greater training stability. GAN training involves a minimax game between a generator and a discriminator \cite{saxena2021generative}, which often leads to mode collapse and training instability, necessitating careful hyperparameter tuning and architectural design.  Other GenAI methods such as Variational Autoencoder (VAE) \cite{pinheiro2021variational},
it is a single-step latent variable model that compresses data into a latent space via an encoder and reconstructs it through a decoder. Its training objective is to maximize the Evidence Lower Bound (ELBO) \cite{alemi2018fixing}. However, due to limitations in single-step reconstruction and the ELBO objective, generated images often appear blurry and struggle to capture high-frequency details. Recent years have witnessed a growing number of DDPM-based applications in fluid dynamics, demonstrating their remarkable potential as a powerful generative framework for complex physical systems.  For instance, Thuerey et al. \cite{liu2024uncertainty} applied an uncertainty-aware DDPM to low Mach number airfoil flows, effectively quantifying predictive confidence. Shu et al. \cite{shu2023physics} used DDPM for super-resolution reconstruction from low-fidelity data, recovering fine-scale turbulent structures. Similarly, Huang et al. \cite{huang2025physics} applied DDPM to airfoil flow field reconstruction and achieved accurate results. Holzschuh et al. \cite{holzschuh2023solving} employed DDPMs for solving inverse problems, and Abaidi \cite{abaidi2025exploring} explored their use in compressible flow prediction. Yang et al. \cite{yang2023denoising} further demonstrated DDPMs' effectiveness in general flow field prediction, establishing them as a robust tool for flow modeling.  Despite these advances, DDPM applications in hypersonic flows remain scarce due to complex flow features such as Shock Wave Boundary Layer Interactions (SWBLI) \cite{gaitonde2013progress,pasquariello2017unsteady}. Thus, this study aims to develop a unified model capable of reconstructing flow fields from incompressible to hypersonic regimes.

Despite the impressive generative fidelity, DDPMs have considerable computational costs due to iterative sampling in high-dimensional pixel space \cite{pandey2022diffusevae}. This process requires a U-Net architecture to sequentially process inputs comprising millions of pixels over hundreds or thousands of diffusion and reverse diffusion steps.  Each step involves expensive operations, particularly intensive memory and attention mechanisms applied to full-resolution feature maps, making training and inference resource-intensive and slow. To overcome this fundamental efficiency bottleneck, Latent Diffusion Models (LDMs) \cite{rombach2022high} have been introduced.  The key innovation lies in performing the diffusion process in a compressed latent space rather than in pixel space. This is achieved through a two-stage framework: first, a pre-trained encoder (typically from a convolutional autoencoder) robustly compresses a high-resolution image  (e.g.,  $512\times512$) into a significantly lower-dimensional latent representation (e.g., $64\times64$). This compression achieves a drastic reduction in the elements of the diffusion model, often by one to two orders of magnitude. All diffusion and denoising operations are then performed in this compact space, drastically reducing the data volume processed by the U-Net and lowering the computational complexity of key operations such as attention mechanisms. This allows the model to focus on learning the underlying semantic and high-level structural features of the data—such as the topology of shock waves, vortex cores, and separation regions in fluid flow fields—rather than spending resources modeling fine-grained pixel-level noise. Finally, a decoder network maps the denoised latent representation back to high-resolution pixel space. This approach greatly improves computational efficiency without compromising generative quality, enabling real-time execution on consumer hardware and broadening the applicability of AI-generated content technologies.

In this study, we leverage both DDPM and LDM to predict temperature distributions on plates with circular or square holes, flow fields around the airfoil, and hypersonic flow fields over a compression ramp. We explore the potential of diffusion models across incompressible and compressible flow regimes, demonstrating their strong generalization. To address the high computational cost of pixel-based diffusion training, we implement a latent-space diffusion framework that maintains competitive accuracy while significantly reducing computational overhead.

The remainder of this paper is organized as follows. Section~\ref{methodology} outlines the methodology, including network architectures and training procedures. Section~\ref{experimetal cases} describes the experimental cases, numerical setups, and data generation methods. Section~\ref{results and analysis} presents and discusses the results on the performance of DDPM and LDM, further demonstrating the predictive capability of LDM in high-dimensional parameter spaces.

%
%
\section{ Methodology }
\label{methodology}
This section outlines the foundational models and methodologies employed in this study. We first describe the principles and architecture of the denoising diffusion probabilistic model, followed by its efficient latent‑space counterpart, the latent diffusion model.  Finally, we present the evaluation metrics adopted to assess the prediction accuracy of both models.
\subsection{Denoising Diffusion Probabilistic Model (DDPM)}
\begin{figure}[htb!]
	\centering
	\includegraphics[trim = 4.5cm 4.3cm 5.5cm 4.5cm, clip,width=\textwidth]{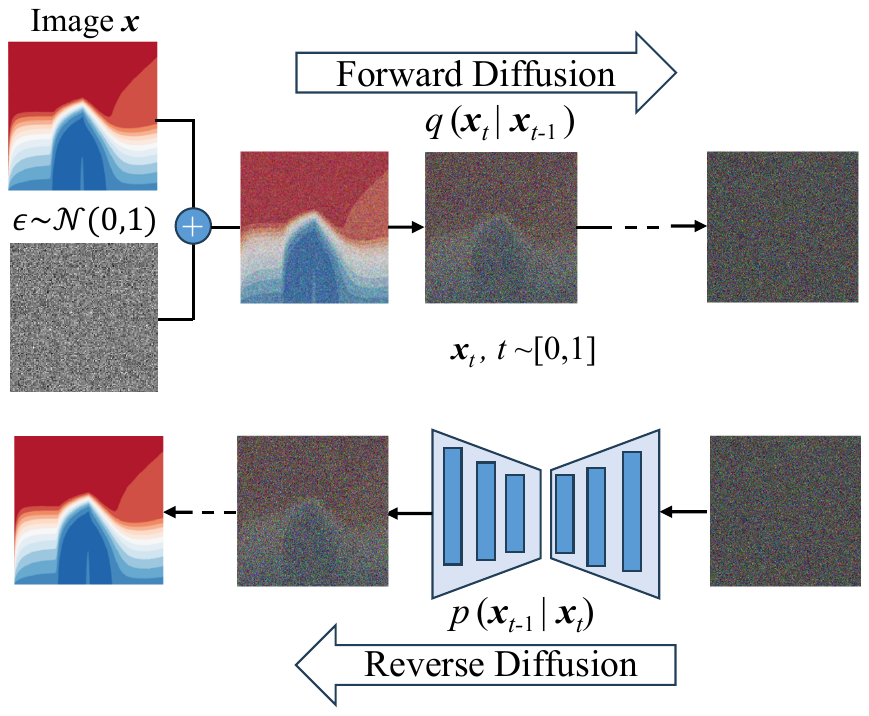}
	\caption{{The architecture of the Denoising Diffusion Probabilistic Model (DDPM).}}
	\label{figs:DDPM}
\end{figure}
Figure~\ref{figs:DDPM} illustrates the architecture of the denoising diffusion probabilistic model adopted in this study. The DDPM consists of two fundamental processes: the forward diffusion process and the reverse diffusion process. The forward process involves progressively adding Gaussian noise to the data until it is completely transformed into pure noise. Specifically, the forward process $q(\boldsymbol{x}_{t}|\boldsymbol{x}_{t-1})$ describes the transition from $\boldsymbol{x}_{t-1}$ to  $\boldsymbol{x}_{t}$ by injecting Gaussian noise at each step. It is formulated as:
\begin{equation}
q(\boldsymbol{x}_{t}|\boldsymbol{x}_{t-1}) = \mathcal{N}(\boldsymbol{x}_{t}|\sqrt{1-\beta_{t} } \boldsymbol{x}_{t-1},\beta_{t}\boldsymbol{I}) ,
\end{equation}
\begin{equation}
\boldsymbol{x}_{t} = \sqrt{\alpha_{t}}\boldsymbol{x}_{t-1}+\sqrt{1-\alpha_{t}}\boldsymbol{\epsilon},
\boldsymbol{\epsilon}\sim \mathcal{N}(0,\boldsymbol{I}),
\end{equation}
where the $\beta_{t}$ denotes the variance schedule controlling the amount of noise, $\boldsymbol{I}$ is the unit tensor, and $\alpha_{t}=1-\beta_{t}$. In this work, we adopt the cosine schedule following \cite{nichol2021improved,liu2024uncertainty}. Additionally, $\boldsymbol{x}_{t}$ can be directly derived from the original image $\boldsymbol{x}_{0}$ as:
\begin{equation}
\boldsymbol{x}_{t} = \sqrt{\overline{ \alpha}_{t}}\boldsymbol{x}_{t-1}+\sqrt{1-\overline{ \alpha}_{t}}\boldsymbol{\epsilon},
\overline{ \alpha}=\prod_{t=1}^{t}\alpha_i,
\end{equation}

The overall forward process can be represented as a Markov chain from $t = 1$ to $t = T$.
 \begin{equation}
q(\boldsymbol{x}_{0:T}) = q(\boldsymbol{x}_{0})\prod_{t=1}^{T}q(\boldsymbol{x}_{t}|\boldsymbol{x}_{t-1}) .
\end{equation}

The reverse diffusion process aims to reconstruct the original data by iteratively denoising from pure noise. This is achieved by training a neural network to predict the noise added at each timestep, which is then gradually subtracted from $\boldsymbol{x}_{t}$ to recover $\boldsymbol{x}_{0}$. The reverse process is defined as:
 \begin{equation}
p(\boldsymbol{x}_{0:T}) = p(\boldsymbol{x}_{T})\prod_{t=1}^{T}p(\boldsymbol{x}_{t-1}|\boldsymbol{x}_{t}) ,
\end{equation}
where
\begin{equation}
p(\boldsymbol{x}_{t-1}|\boldsymbol{x}_{t}) = \mathcal{N}(\boldsymbol{x}_{t-1},\mu(\boldsymbol{x}_{t},t),\sum(\boldsymbol{x}_{t},t)) ,
\end{equation}
and $\mu(\boldsymbol{x}_{t},t)$ is given by:
\begin{equation}
\mu(\boldsymbol{x}_{t},t) = \frac{1}{\sqrt{ \alpha_{t}}}(\boldsymbol{x}_{t}-\frac{\beta_{t}}{\sqrt{ 1-\overline{\alpha_{t}}}}\boldsymbol{\epsilon}(\boldsymbol{x}_{t},t)).
\end{equation}
Here, $\boldsymbol{\epsilon}(\boldsymbol{x}_{t},t)$ denotes the noise predicted by the neural network based on $(\boldsymbol{x}_{t},t)$, while $\mu(\boldsymbol{x}_{t},t)$ represents the deterministic denoising direction derived from this prediction.

\begin{figure}
	\centering
	\begin{subfigure}[b]{0.99\textwidth}
		\centering
		\includegraphics[trim = 0cm 1cm 0cm 0cm, clip,width=0.99\textwidth]{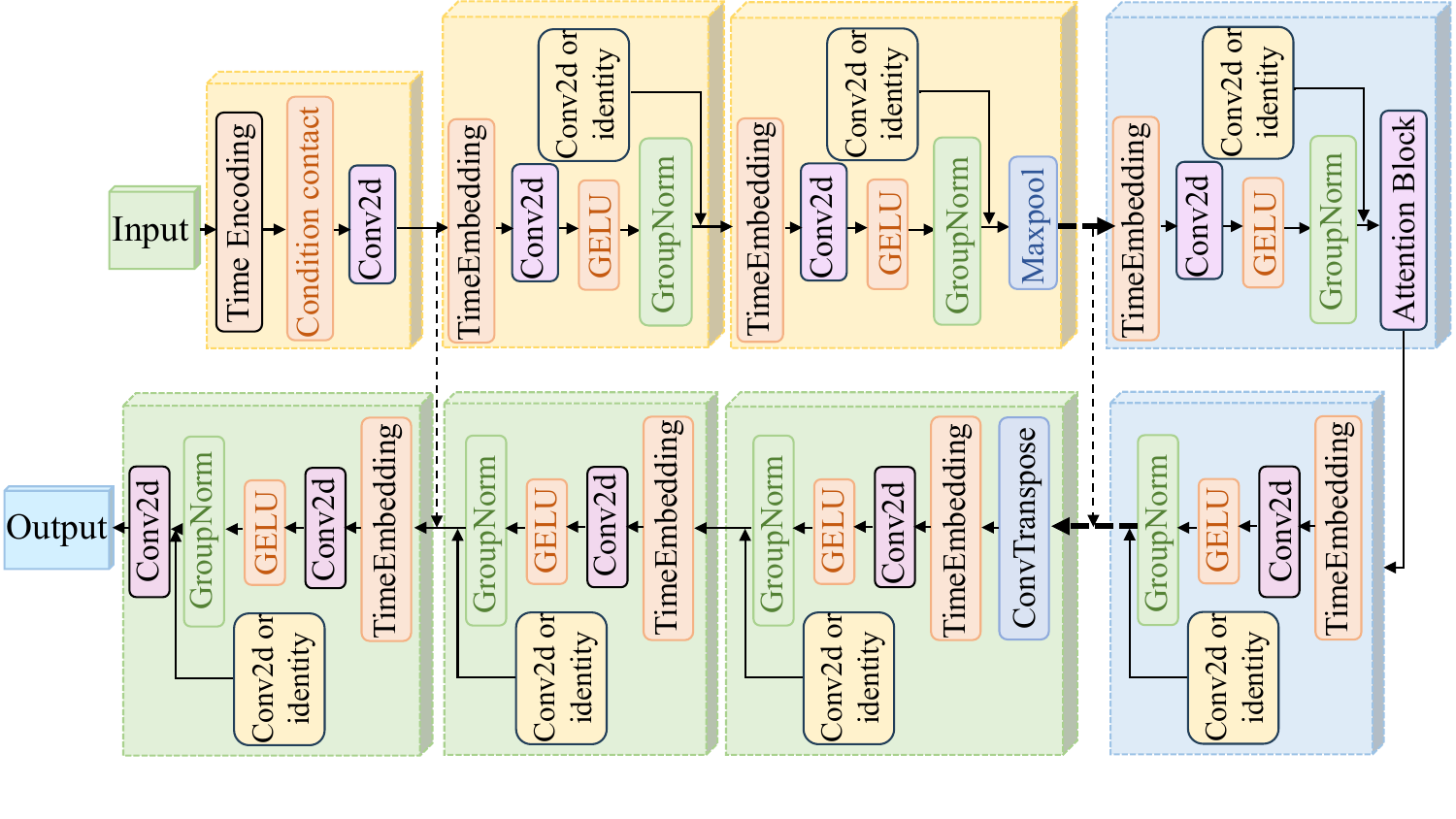}
		\caption{}
		\label{figs:re-constructed(a)}
	\end{subfigure}
        \begin{subfigure}[b]{0.95\textwidth}
		\centering
		\includegraphics[trim = 2cm 4cm 2cm 0cm, clip, width=\textwidth]{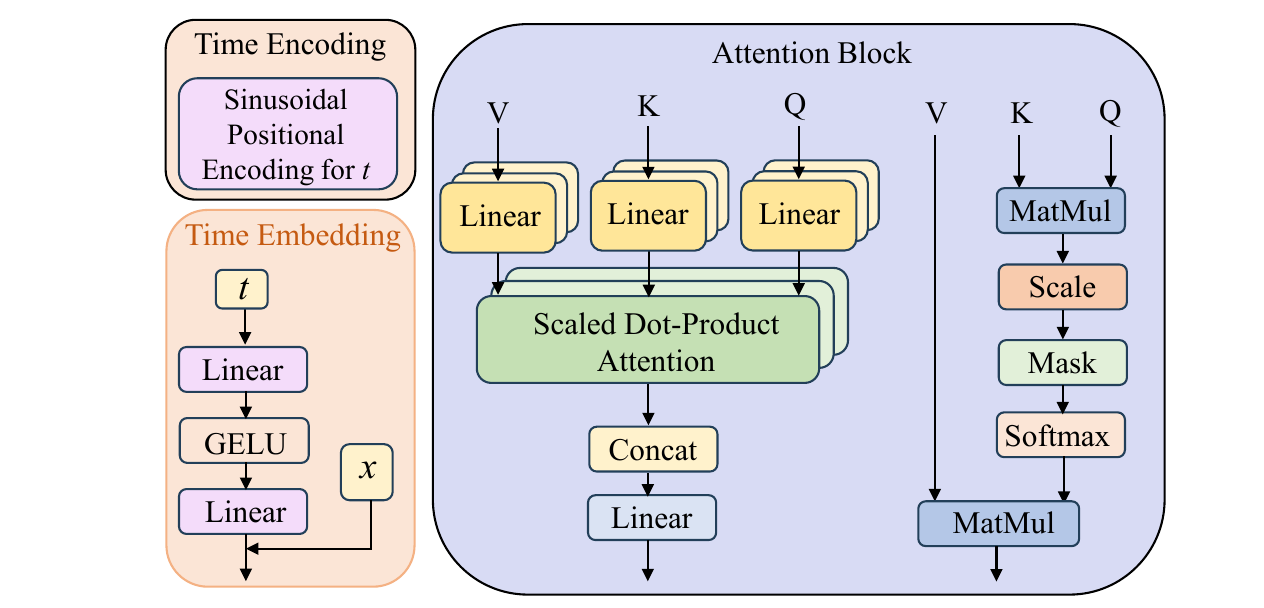}
		\caption{}
		\label{figs:dddd}
	\end{subfigure}
	\caption{Details of the DDPM. (a): The network structure; Dashed lines represent skip connections; Bold dash lines denote that not all layers are displayed; (b): Detail of each block. }
    \label{figs:DDPM details}
\end{figure}
Figure~\ref{figs:DDPM details} illustrates details of each block for the DDPM build on the U-Net architecture employing the classical encoder-decoder structure for conditional image generation tasks. The network progressively extracts features through downsampling path utilizing maxpooling for dimensionality reduction. Following the bottleneck layer, the resolution is restored via upsampling path, with feature information from different levels fused through skip connections. The core of this model lies in the deep integration of temporal step $t$ and conditional information. The original time scalar $t$ first undergoes a feature transformation via sinusoidal positional encoding function. This sinusoidal encoding is then fed into a time embedding network, and then coupling with spatial features.

A multi-head attention mechanism is also incorporated within the DDPM architecture. 
In this mechanism, the input is projected into Query ($Q$), Key ($K$), and Value ($V$) matrices through linear transformations. The output of each attention head $head_{i}$ is computed as:
\begin{equation}
{head}_i = {softmax}\left(\frac{Q_i K_i^T}{\sqrt{d_k}}\right) V_i
\end{equation}
where $Q_i K_i^T$ computes attention scores, scaled by $\sqrt{d_k}$ -- the dimension of key vectors in each attention head -- to ensure numerical stability. The Softmax function then converts these scores into probabilistic weights for weighted aggregation of the value vectors. The outputs of all attention heads are concatenated as follows:

\begin{equation}
{MultiHead}(Q,K,V) = {Concat}({head}_1, ..., {head}_h)W^O,
\end{equation}
where  $W^O$ is a final linear projection that integrates the multi-head features into the output dimension.

In this study, we utilize a conditional DDPM framework to enable conditional sampling. The conditioning variable $\boldsymbol{y}_{con}$  is integrated into the network via $\boldsymbol{\epsilon}(\boldsymbol{x}_{t},t,\boldsymbol{y}_{con})$, and the corresponding training loss is defined as:
\begin{equation}
	  \mathcal{L}_{NN} = \mathbb{E}_{\boldsymbol{x}} \left[ || \boldsymbol{\epsilon} - \boldsymbol{\epsilon}(\boldsymbol{x}_{t},t,\boldsymbol{y}_{con}) ||^2\right] 
	\label{momentum-equation}.
\end{equation}
\subsection{Latent Diffusion Model (LDM)}
\begin{figure}[htb!]
	\centering
	\includegraphics[trim = 0cm 0.5cm 0cm 1cm, clip,width=0.9\textwidth]{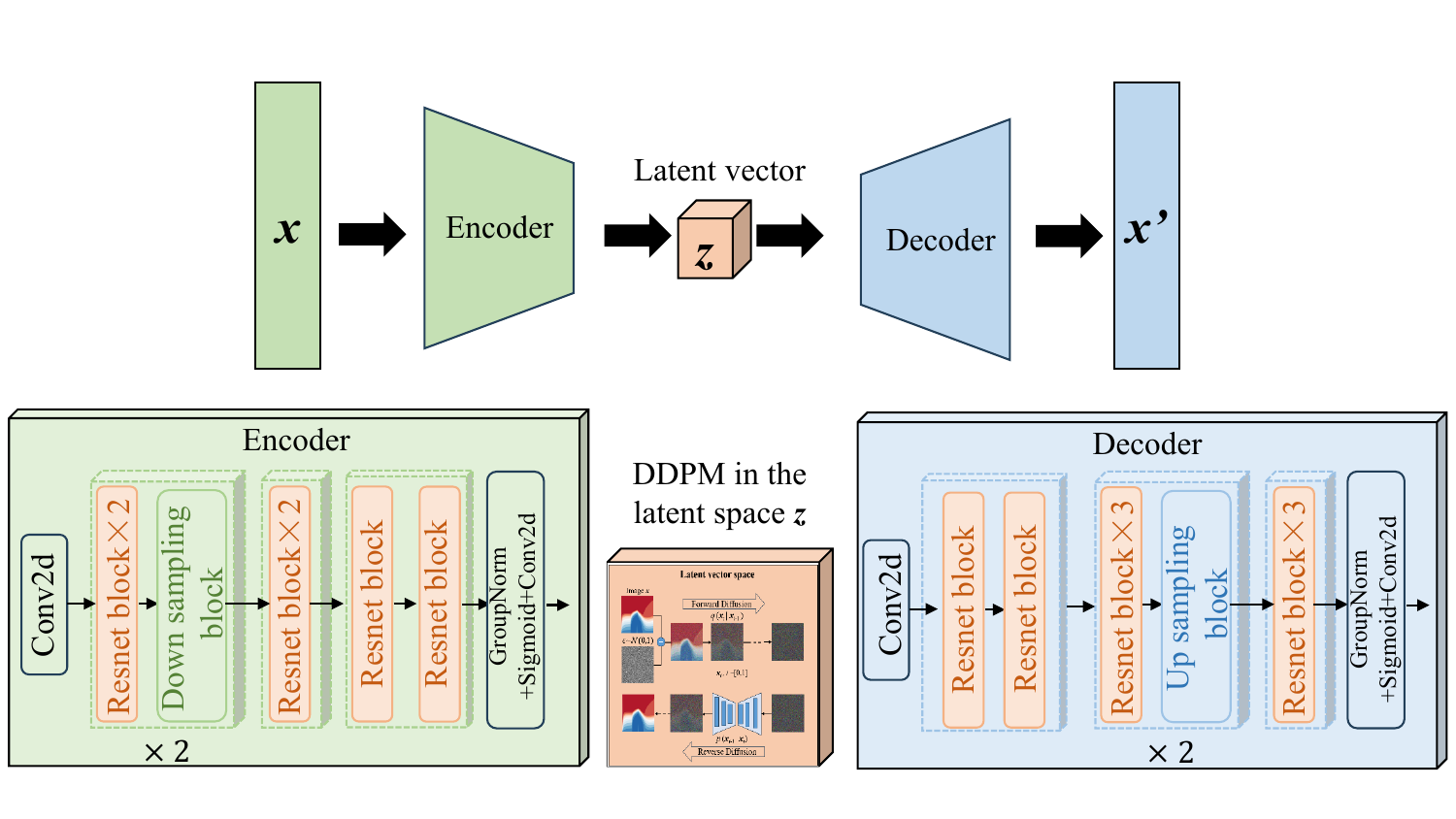}
	\caption{{The architecture of the Latent Diffusion Model (LDM) with details of each block.}}
	\label{figs:LDM}
\end{figure}
Figure~\ref{figs:LDM} shows the overall architecture of the LDM and its key components. The model first trains an encoder to obtain a compact latent space that captures efficient low‑dimensional representations of the input data. As highlighted in the bottom green box, the encoder consists of multiple building blocks, including downsampling modules and several ResNetBlocks. Subsequently, a decoder (bottom blue box) reconstructs the data from the latent space back to the original parameter space; it similarly consists of multiple building blocks, including upsampling modules and several ResNetBlocks.

Each ResNetBlock follows a residual‑learning structure, containing the sequential operations: Group Normalization (GroupNorm), a SiLU activation, and a convolutional layer. The transformed features are added to the original input through a residual connection, which facilitates stable gradient flow throughout the network. Notably, the number of input channels in each ResNetBlock remains the same as the number of output channels.

The diffusion process is performed in this latent space, with the training loss for the LDM defined as:
\begin{equation}
	  \mathcal{L}_{LDM} = \mathbb{E}_{z} \left[ || \boldsymbol{\epsilon} - \boldsymbol{\epsilon}(\boldsymbol{z}_t,t) ||^2\right] 
	\label{momentum-equation},
\end{equation}
where $z$ is the representation in the latent space.
\subsection{Evaluation Metrics}
To validate the prediction accuracy of the proposed model, the error evaluation criteria are defined as follows:
\begin{equation}
	  \bm{\zeta}  =  ||\mathbf{Q}-\hat{\mathbf{Q}}||_{1} 
	\label{momentum-equation},
\end{equation}
where $\bm{\zeta}$  represents the overall magnitude error across the domain between the ground truth $\mathbf{Q}$ and the predicted flow field $\hat{\mathbf{Q}}$.  The training process is performed on an NVIDIA GeForce RTX 4090 GPU using the AdamW optimizer \cite{loshchilov2017decoupled}.

\section{ Experimental cases }
\label{experimetal cases}
\begin{table}[h]
\centering
\caption{Summary of three physical fields and corresponding methods.}
\label{tab:case_summary}
\scriptsize
\begin{tabular}{|p{3.3cm}|p{3.5cm}|p{3cm}|p{2cm}|}
\hline
\textbf{Case Description} & \textbf{Physical Characteristics} & \textbf{Methodological Approach} & \textbf{Differentiating Factors} \\
\hline
\hline
\textbf{Temperature distribution for plates with holes}: \textbf{Purpose}: Baseline validation & 
Steady-state heat conduction; 
linear physics; 
constant material properties &
Mask-based input representation &
\textbf{Simplicity}: Linear governing equation \\
\hline
\hline
\textbf{Airfoil flow field}: \textbf{Purpose}: Demonstrates the flow features of boundary layer and pressure gradient & 
Steady incompressible flow;  
boundary layer development; 
pressure-velocity coupling &
Mask-based boundary conditions &
\textbf{Complexity}: Nonlinear governing equations\\

\hline
\hline
\textbf{Hypersonic compression corner}:  \textbf{Purpose}: Demonstrates model generalization& 
Compressible flow; 
shock wave boundary layer interaction; 
high temperature effects; 
severe gradients &
Implicit via coordinate transformation (conditions by mesh transformation) &
\textbf{Extreme conditions}: Shock waves, compressibility \\
\hline
\end{tabular}
\label{tab:three methods}
\end{table}
Table~\ref{tab:three methods} summarizes the thermal/flow fields and corresponding methods. This study systematically verifies the universality of the proposed model through three progressive computational examples. Firstly, the temperature field of plates with square or circular holes serves as a linear benchmark problem, validating the model's fundamental solution capability based on the mask-based learning paradigm. Secondly, the flow field around an airfoil incorporates the nonlinearity of the Navier-Stokes equations and pressure-velocity coupling, examining the model's handling of complex incompressible viscous flows. Finally, hypersonic flow around a compression corner extends the analysis to compressible flow fields, incorporating extreme physical phenomena such as shock wave boundary layer interactions. Through coordinate transformation methods, the model maps non-uniform physical grids into uniform computational grids, demonstrating its ability to transcend traditional uniform grid limitations and generalize to complex computational domains with steep gradients encountered in real-world engineering applications. These three case studies form a comprehensive validation framework across three dimensions—physical nonlinearity, compressibility, and mesh adaptability—highlighting the potential value of this methodology from fundamental validation to engineering applicability.

\subsection{Plate temperature distributions}

\begin{figure}[htb!]
	\centering
	\includegraphics[trim = 3cm 4.4cm 3cm 2.3cm, clip,width=0.7\textwidth]{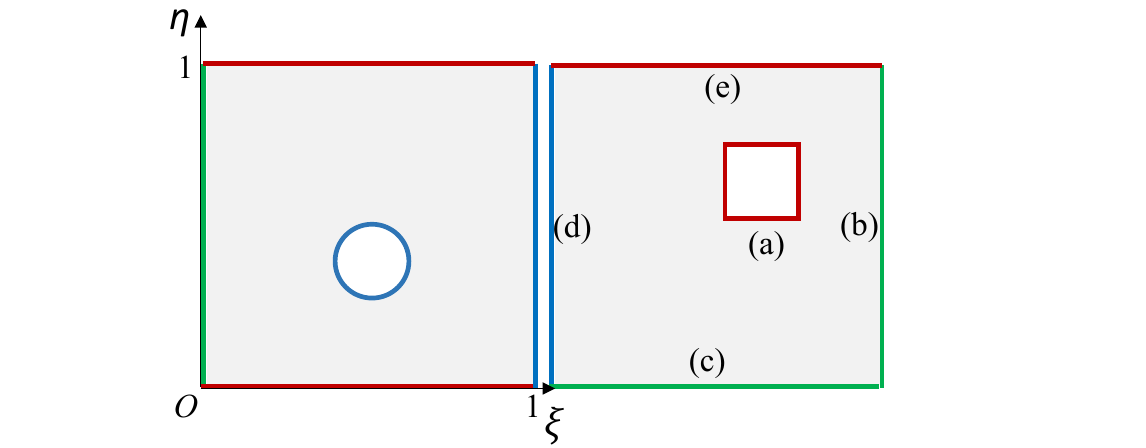}
    	\caption{{The schematic diagram for plates with holes. (a)-(e) represent 5 boundary conditions, blue corresponds to a temperature of 0, green to 0.5, and red to 1.}}
	\label{figs:plateboundary}
\end{figure}

We first evaluate the neural network model for predicting plate temperature distributions across varying geometries (plates with square or circular holes at different locations) and boundary temperatures ($T_{\text{boundary}} = 0/0.5/1$). As shown in Figure~\ref{figs:plateboundary}, hole positions are varied over nine locations: $(\xi, \eta) = (0.5 \pm 0.125, 0.5 \pm 0.125)$, with both circle diameter and square side length fixed at $0.25$. The five boundaries have three possible temperatures (blue: 0; green: 0.5; red: 1), thus, the dataset contains $3^5 = 243$ samples with different boundary conditions per geometry.

For the DDPM, the input is denoted as $\mathcal{G} \in \mathbb{R}^{C_{\text{in}} \times W \times H} = \mathbb{R}^{2 \times 128 \times 128}$, where $C_{\text{in}}$ is the number of input channels, $W$ and $H$ represent the width and height of the input image, respectively. The output is represented as $\mathcal{Q} \in \mathbb{R}^{C_{\text{out}} \times W \times H} = \mathbb{R}^{1 \times 128 \times 128}$, where $C_{\text{out}}$ is the number of output channels, corresponding to the temperature $T$ at each spatial point. According to Fourier's law, under the conditions of constant thermal conductivity and no internal heat generation, the governing equation for heat conduction reduces to the two-dimensional Laplace equation, which is solved numerically using OpenFOAM to generate the temperature distribution data for the plate.

\begin{equation}
	  \nabla^{2}T = \frac{\partial^2 T}{\partial x^{2}} + \frac{\partial^2 T}{\partial y^{2}} =0
	\label{momentum-equation}.
\end{equation}

For the LDM, both the input $\mathcal{G}$ and output $\mathcal{Q}$ are compressed into the latent space with dimensions $\mathbb{R}^{2 \times 32 \times 32}$ and $\mathbb{R}^{1 \times 32 \times 32}$, respectively. By operating in this reduced space, the image dimensions are decreased by a factor of 16, substantially lowering the computational cost of the diffusion process. In this study, 220 plate cases are used for training, and 23 separate cases are held out for testing.

\subsection{Incompressible airfoil flow fields}
The model is further applied to predict incompressible two‑dimensional airfoil flows governed by the Reynolds‑averaged Navier–Stokes (RANS) equations. We adopt the public dataset released by Thuerey et al. \cite{thuerey2020deep}, which was generated with OpenFOAM using the Spalart–Allmaras turbulence model. At the airfoil surface, the triangular mesh has an average edge length of 1/200 in order to resolve the flow boundary layer \cite{thuerey2020deep}. Each computational case is defined by a distinct airfoil geometry selected from the UIUC database, together with freestream velocity, Reynolds number $Re$ ranging from $5\times10^5$ to $5\times10^6$, and angle of attack $\alpha'$ varying between $-22.5^\circ$ and $22.5^\circ$.

The DDPM input $\mathcal{G} \in \mathbb{R}^{3\times128\times128}$ consists of a binary mask encoding the airfoil shape and the $x$‑ and $y$‑components of the freestream velocity.  The output $\mathcal{Q} \in \mathbb{R}^{C_{out}\times{W}\times{H}}=\mathbb{R}^{3\times128\times128} $ comprises the velocity components $u$ and $v$, as well as the pressure field $p$. All variables are normalized to the range [0, 1] using min–max scaling. For the LDM, both the input $\mathcal{G}$ and output $\mathcal{Q}$ are mapped to a latent space with dimensions of $\mathbb{R}^{3 \times 32 \times 32}$ and $\mathbb{R}^{3 \times 32 \times 32}$, respectively. This compression reduces the image dimensions by a factor of 16, thereby significantly lowering the computational cost of the subsequent diffusion process. For this study, 160 cases are used for training, and 11 cases are reserved for testing.
\subsection{Hypersonic flow over compression ramp}
\begin{figure}[htb!]
	\centering
	\includegraphics[trim = 3cm 5cm 5cm 4cm, clip,width=0.8\textwidth]{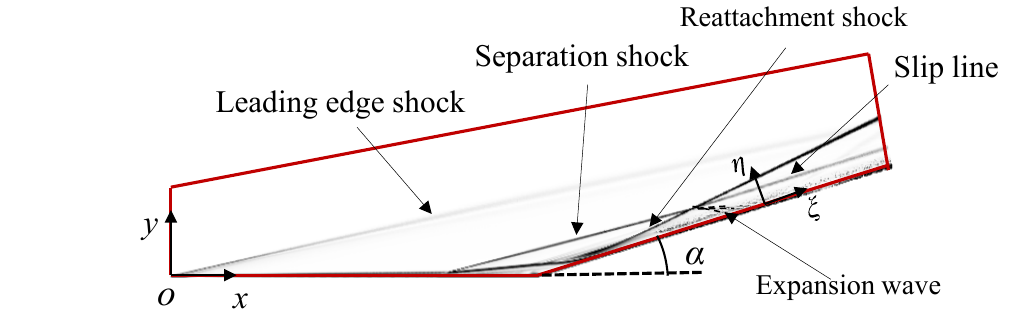}
	\caption{{The hypersonic flow structure over a compression ramp with computational domain in red lines.}}
	\label{figs:ramp1}
\end{figure}

Finally, the proposed model is evaluated by predicting the hypersonic flow fields over a compression ramp. Figure~\ref{figs:ramp1} illustrates the compression ramp configuration, with the computational domain outlined in red.  Based on experiments conducted in the Hong Kong Polytechnic University \cite{zhao2024investigation}, the geometry consists of a flat plate ($L$ = 100 mm) followed by a 100 mm adjustable-angle ramp, a configuration widely used in previous numerical studies \cite{hao2021occurrence,cao2023stability,hao2023response}. A Cartesian coordinate system $(x,y)$ is defined with its origin at the leading edge of the plate. As shown in Figure~\ref{figs:ramp1}, the flow exhibits an Edney type VI shock interaction \cite{edney1968effects}, initiated by a viscous‑induced leading‑edge shock.  Subsequent flow deflection establishes an adverse pressure gradient that triggers separation near the corner \cite{hao2021occurrence}. The resulting separation zone gives rise to separation and reattachment shocks, whose interaction generates a slip line and an expansion wave.

The flow conditions are specified as follows: Mach number $Ma$ ranges from 2 to 9.4, Reynolds number $Re$ from $3.36\times 10^6$ to $5.04\times 10^6$, freestream temperature $ T_{\infty}$ from 100 K to 150 K, ramp angle $\alpha$ from  $9^\circ$ to $18^\circ$, and wall temperature ratio $ T_w /T_0$ from 0.036 to 0.86. Here, $ T_w$ denotes the wall temperature and $ T_0$ represents the  total temperature of the freestream flow. For the computational study, a set of 112 distinct combinations are generated employing Latin Hypercube Sampling (LHS) \cite{shields2016generalization}. The hypersonic compression ramp flow fields are computed via Direct Numerical Simulation (DNS) with the in-house multi-block parallel finite-volume solver, PHAROS \cite{hao2016numerical,hao2020hypersonic}. 
 The compressible Navier-Stokes equations for a calorically perfect gas in 2D is written in the following conservation form:
\begin{equation}
	  \frac{\partial \textbf{\emph{U}}}{\partial \mathit{t}} + \frac{\partial \textbf{\emph{F}}}{\partial \mathit{x}}+\frac{\partial \textbf{\emph{G}}}{\partial \mathit{y}} =\frac{\partial \boldsymbol{F}_{v}}{\partial \mathit{x}}+\frac{\partial \boldsymbol{G}_{v}}{\partial \mathit{y}},
	\label{momentum-equation}
\end{equation}
where  $\textbf{\emph{U}} = (\rho,\rho u,\rho v,\rho e)^T$ is the vector of conservative variables, $\rho$ stands for density, $u$ and $v$ are the flow velocities in the $x$ and $y$  directions, respectively, and $e$ is the total energy per unit mass. $\textbf{\emph{F}}$ and $\textbf{\emph{G}}$  represent the inviscid fluxes,  while $\boldsymbol{F}_{v}$ and $\boldsymbol{G}_{v}$ denote the viscous fluxes.
The inviscid fluxes are discretized using the modified Steger-Warming central-difference scheme \cite{maccormack2014numerical}, and the viscous fluxes are evaluated with a second-order central-difference method. For the boundary conditions: Free-stream conditions are imposed on the upper and left boundaries; An extrapolation outflow condition is applied at the exit and a no-slip condition is enforced on the wall surface. The wall pressure coefficient $C_p$ is defined as:
 \begin{equation}
		{C}_{p}=\frac{{p}_{w}}{0.5{\rho }_{\infty }{u}_{\infty }^{2}},
	\end{equation}
 where $p_w$ is the wall pressure, ${\rho }_{\infty }$ and ${u}_{\infty }$  are the freestream density and velocity, respectively.
Based on the grid dependence analysis, PHAROS simulations employ a computational grid comprising 1200 streamwise and 400 wall-normal points. This configuration, featuring a surface-normal spacing of $  1\times  10^{-7} $ m, aligns with previous studies \cite{hao2021occurrence} and has been experimentally validated.

A coordinate transformation is applied from the Cartesian system $(x,y)$ to the curvilinear coordinate system $ (\xi ,\eta ) $, as shown in Figure~\ref{figs:ramp1}. Based on the coordinate transformation above, the input features to the neural network comprise the following parameters: $Ma$, $Re$, $T_{\infty}$, $\alpha$, $T_w/T_0$, along with six coordinate transformation parameters - the compression ramp geometry coordinates ($x_0$, $y_0$) and the four components ($\xi_x$, $\xi_y$, $\eta_x$, $\eta_y$) from the transformation process. The detailed equations are as follows:
\begin{equation}
	  \begin{bmatrix}x\\ y
\end{bmatrix}_{{i},{j}}=\begin{bmatrix} x_0\\ y_0
\end{bmatrix}_{{i},0} +\int\begin{bmatrix}
 dx\\ dy
\end{bmatrix}
	\label{momentum-equation},
\end{equation}
where $i$ and $j$ are the grid nodes in the streamwise and normal directions, respectively. And the coordinates in the curvilinear system can be normalized by the total number of grid nodes in the streamwise direction $i_{max}$ and normal direction $j_{max}$:
\begin{equation}
	  \xi=\frac{(i-1)}{(i_{max} -1)}, \eta=\frac{(j-1)}{(j_{max} -1)}.
\end{equation}
The transformation matrix $T$ can be denoted by:
\begin{equation}
	   {T}=\begin{bmatrix} {\xi}_x  & {\xi}_y\\ \eta_x & \eta_y
\end{bmatrix} 
	\label{momentum-equation}.
\end{equation}

The original high-resolution input for the compression ramp problem is $\mathcal{G} \in \mathbb{R}^{11\times1200\times400} $. The original output $\mathcal{Q} \in \mathbb{R}^{C_{out}\times{W}\times{H}}=\mathbb{R}^{4\times1200\times400} $ comprises the velocity components $u$ and $v$, as well as the pressure field $p$ and density $\rho$. All variables are normalized to the range [0, 1] using min–max scaling, and the predicted flow fields are de-normalized to the original parameter space for better comparison.

Firstly, to enable a fair comparison between DDPM and LDM architectures, we establish a common benchmark at a lower resolution. The original data $\mathbf{X} \in \mathbb{R}^{15 \times 1200 \times 400}$, with the 15 dimensions representing 11 input channels and 4 output channels, are further downsampled to $\mathbf{Y} \in \mathbb{R}^{15\times150\times50}$ using $8\times8$ average pooling:
\begin{equation}
	  \mathbf{Y}[i,j,c] = \frac{1}{64} \sum_{p=0}^{7}\sum_{q=0}^{7} \mathbf{X}[8i+p,8j+q,c] 
	\label{momentum-equation},
\end{equation}
where $i = 0,1,\dots,149$ (output width index), $j = 0,1,\dots,49$ (output height index), $c = 0,1,\dots,14$ (channel index). The model input $\mathcal{G} \in \mathbb{R}^{11\times150\times50} $ contains the freestream flow and geometry conditions. The output $\mathcal{Q} \in \mathbb{R}^{C_{\text{out}}\times W\times H} = \mathbb{R}^{4\times150\times50}$ includes the velocity components $u$ and $v$, pressure $p$, and density $\rho$. The LDM framework trains on data downsampled from $\mathbb{R}^{150\times 50}$ to $\mathbb{R}^{30\times 10}$ in spatial resolution with a scale factor of 25. 

Then, the original high-resolution is applied in the LDM framework, the spatial resolution of the original data is reduced from $\mathbb{R}^{1200\times 400}$ to $\mathbb{R}^{300\times 100}$ for training, thereby effectively lowering computational costs. The model is trained on 89 samples, with 13 used for validation and 10 reserved for testing.

\section{ Results and analysis }
\label{results and analysis}
\subsection{Predictions of plate temperature distribution}



\begin{table}[htbp]
\centering
\caption{The average errors of DDPM and LDM for plate temperature distributions across all test cases.}
\scriptsize 
\setlength{\tabcolsep}{3pt} 
\begin{tabular}{c*{6}{c}}
\toprule
\multirow{2}{*}{Model} & \multicolumn{3}{c}{Circular holes $\bm{\zeta}_{T}$} & \multicolumn{3}{c}{Square holes $\bm{\zeta}_{T}$} \\
\cmidrule(lr){2-4} \cmidrule(lr){5-7}
& (0.5,0.5) & (0.5,0.375) & (0.625,0.625) & (0.5,0.5) & (0.375,0.5) & (0.625,0.625) \\
\midrule
DDPM & 0.013471 & 0.037264 & {\color{blue}0.002147} & 0.048297 & {\color{blue}0.004233}& 0.038570 \\
LDM & {\color{blue}0.013267} & {\color{blue}0.012054} & 0.007686 & {\color{blue}0.025625} & 0.005758 & {\color{blue}0.013074} \\
\bottomrule
\end{tabular}
\label{tab:angular_performance1}
\end{table}

The models are first evaluated on the temperature distribution of the plate with holes. Table~\ref{tab:angular_performance1} illustrates the prediction errors of DDPM and LDM for all test cases with different geometries and boundary conditions, with the lower error for each condition marked in blue. Overall, the LDM demonstrates accuracy that is either superior to or comparable with the DDPM. 

 For the plate with a central circular hole, LDM achieves a lower overall prediction error than DDPM, reducing the error from 0.013471 to 0.013267 across all test cases. This result indicates that encoding the image into a latent space for the diffusion process preserves predictive performance comparable to the standard DDPM, while even providing a slight error reduction. The marginal improvement may also be attributed to the inherent stochasticity in the diffusion model's noise‑sampling process, which can introduce beneficial variability that occasionally enhances the predictions.
\begin{figure}[htb!]
	\centering
	\includegraphics[trim = 2cm 0cm 1.5cm 0.2cm, clip,width=\textwidth]{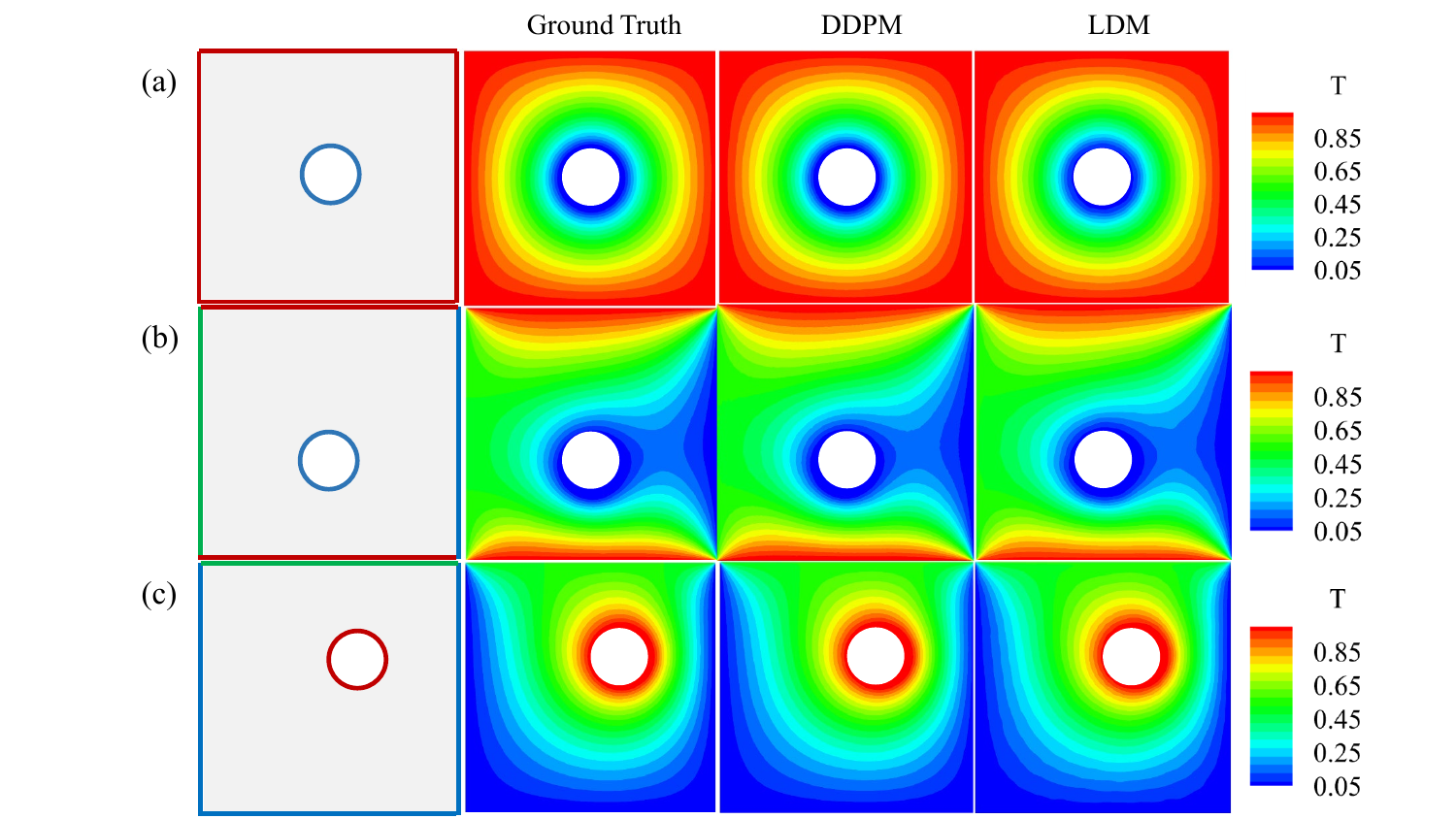}
	\caption{{Results of plate temperature distributions with circular holes. The color in the first column represents the boundary temperature values: blue corresponds to a temperature of 0, green to 0.5, and red to 1. Hole positions: (a), (0.5,0.5); (b), (0.5,0.375); (c), (0.625,0.625).}}
	\label{figs:circularplates}
\end{figure}
\begin{figure}[htb!]
	\centering
	\includegraphics[trim = 2cm 0.1cm 2cm 0cm, clip,width=\textwidth]{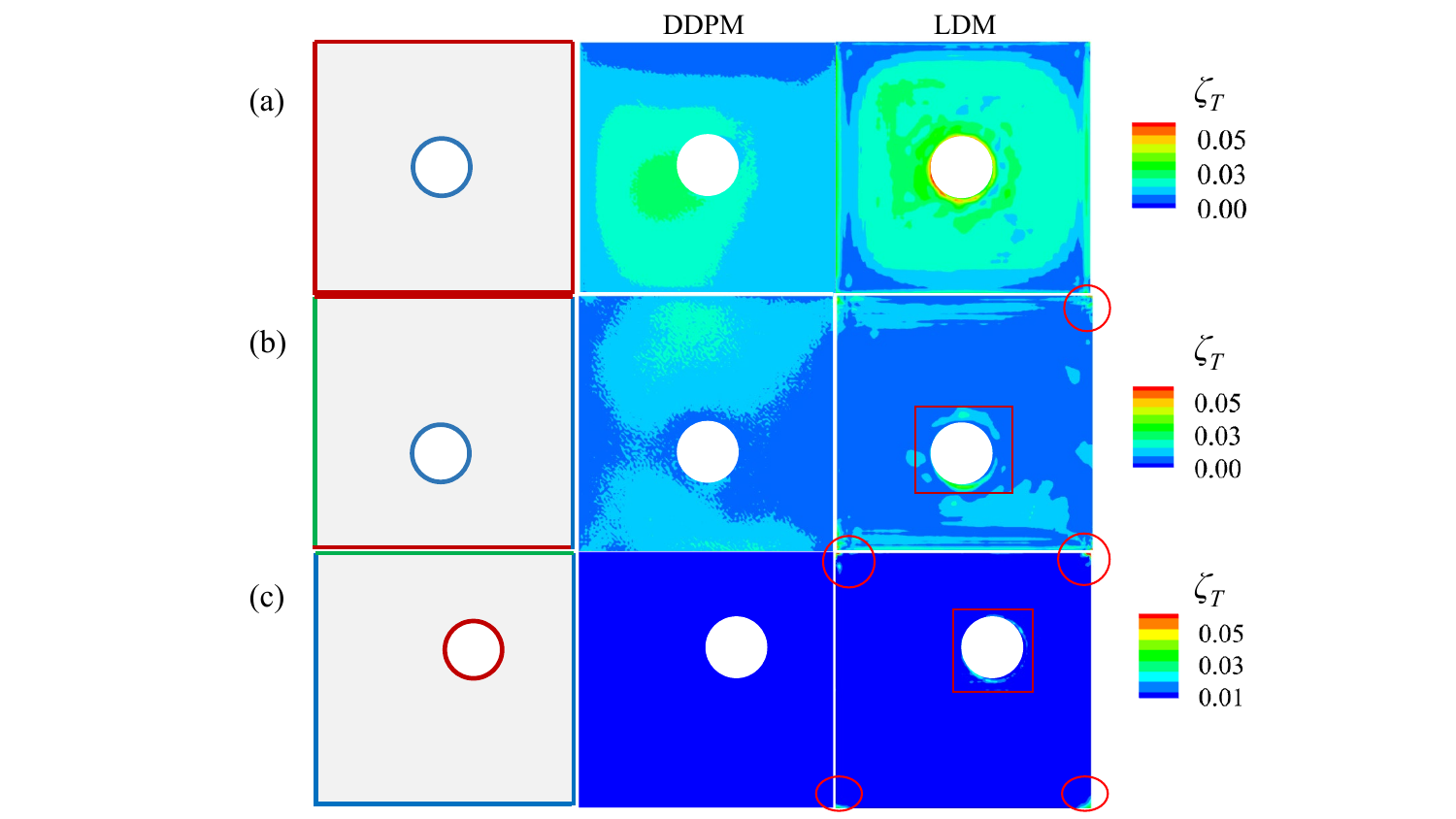}
	\caption{{Results of errors in plate temperature distributions with circular holes. The color in the first column represents the boundary temperature values: blue corresponds to a temperature of 0, green to 0.5, and red to 1. Hole positions: (a), (0.5,0.5); (b), (0.5,0.375); (c), (0.625,0.625).}}
	\label{figs:circularplateserror}
\end{figure}

Figure~\ref{figs:circularplates} compares the temperature distributions of the ground truth with those predicted by DDPM and LDM for plates with circular holes at varying locations. Each model is evaluated on three geometric configurations: the colors in the first column denote different temperature boundaries, while the remaining columns present the ground truth and corresponding predictions. Groups (a), (b) and (c) at three rows present the distinct geometries with circular holes at different positions.
 Both models achieve high accuracy and closely align with the ground truth. As shown in Figure~\ref{figs:circularplateserror}, for groups (a) and (b) by DDPM, the speckled or granular error patterns observed are most likely caused by noise inherent in its per-pixel denoising process. Although LDM delivers performance comparable to DDPM, it exhibits slightly reduced smoothness along the interior hole boundary.  The error contours in Figure~\ref{figs:circularplateserror} reveal that LDM errors are predominantly concentrated around the internal hole and domain edges, which are marked by red boxes and circles.  This concentration of error mainly results from the compression–reconstruction process inherent to the latent diffusion model. Encoding the flow field parameters into a lower‑dimensional latent space and subsequently decoding them back to the original spatial dimension can smooth fine‑scale geometric features along internal boundaries, thereby localizing errors near the circle. However, from a global perspective, these errors remain within an acceptable range, as overall performance metrics are not significantly affected. 

\begin{figure}[htb!]
	\centering
	\includegraphics[trim = 0cm 0cm 0cm 0.2cm, clip,width=\textwidth]{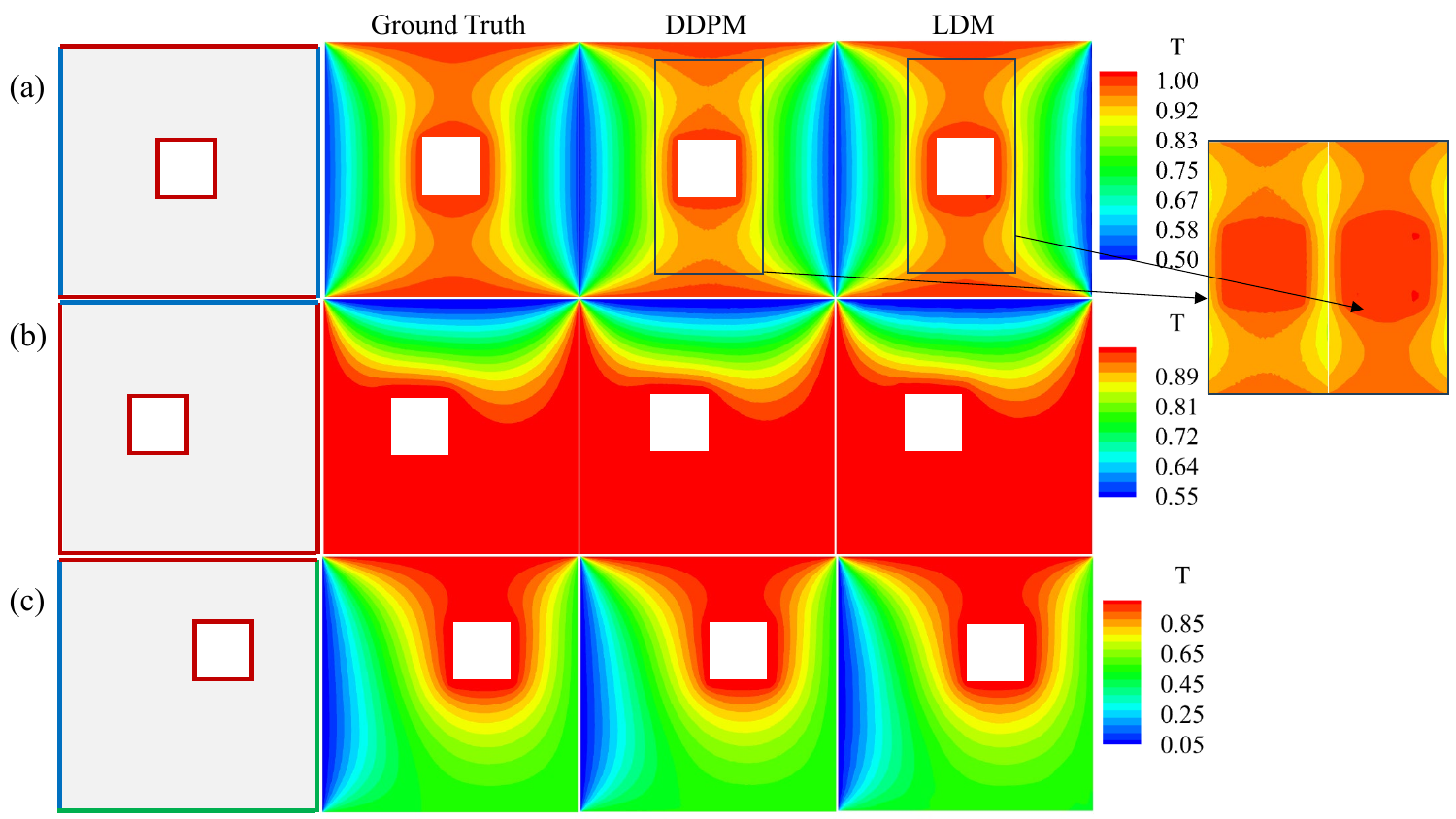}
	\caption{{Results of plate temperature distributions with square holes. The color in the first column represents the boundary temperature values: blue corresponds to a temperature of 0, green to 0.5, and red to 1. Hole positions: (a), (0.5,0.5); (b), (0.375,0.5); (c), (0.625,0.625).}}
	\label{figs:squareplates}
\end{figure}
\begin{figure}[htb!]
	\centering
	\includegraphics[trim = 2cm 0cm 2cm 0cm, clip,width=\textwidth]{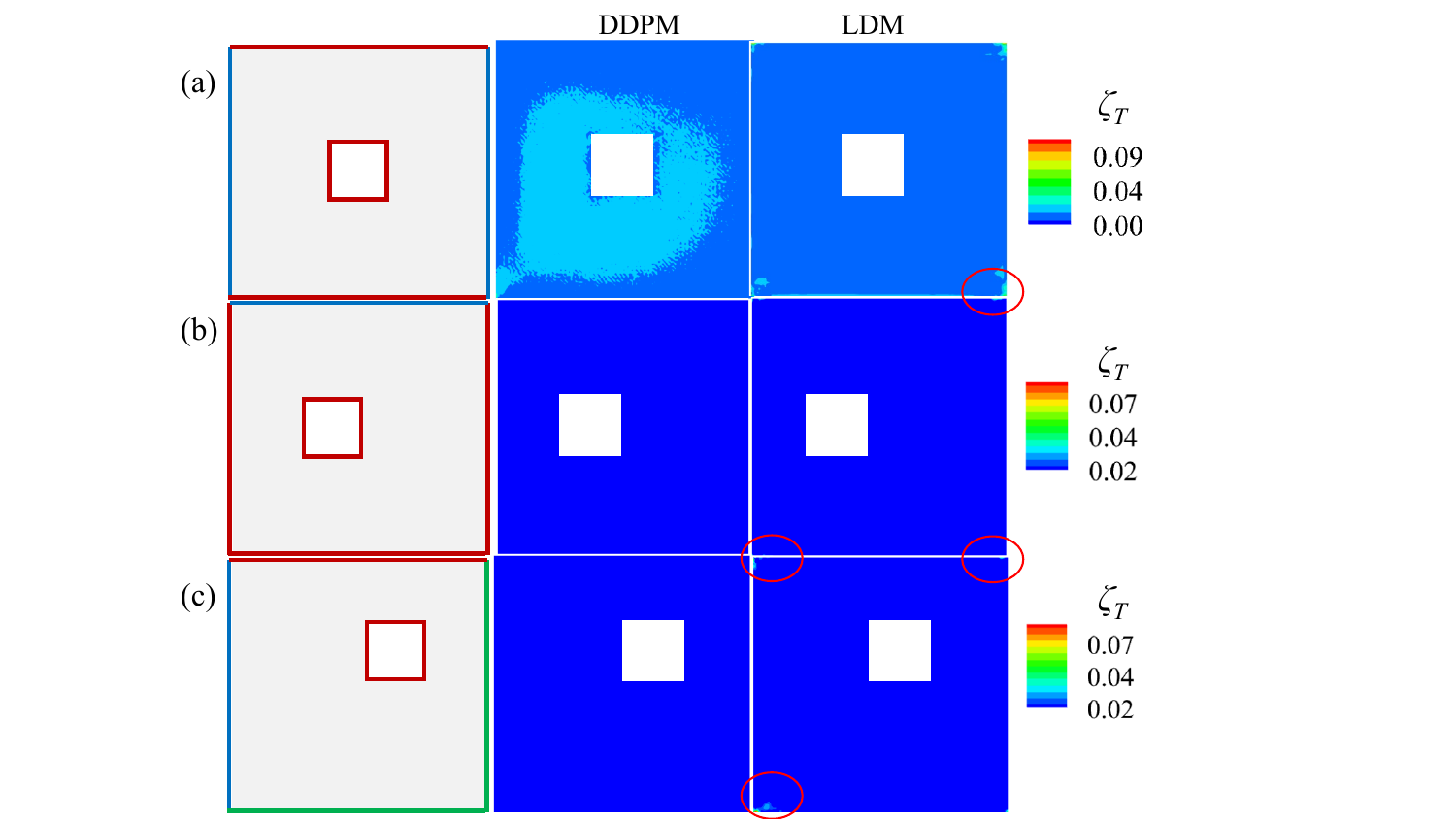}
	\caption{{Results of errors in plate temperature distributions with square holes. The color in the first column represents the boundary temperature values: blue corresponds to a temperature of 0, green to 0.5, and red to 1. Hole positions: (a), (0.5,0.5); (b), (0.375,0.5); (c), (0.625,0.625).}}
	\label{figs:squareplateserror}
\end{figure}

\begin{figure}[htb!]
	\centering
	\includegraphics[trim = 3cm 2.5cm 3cm 1.5cm, clip,width=0.7\textwidth]{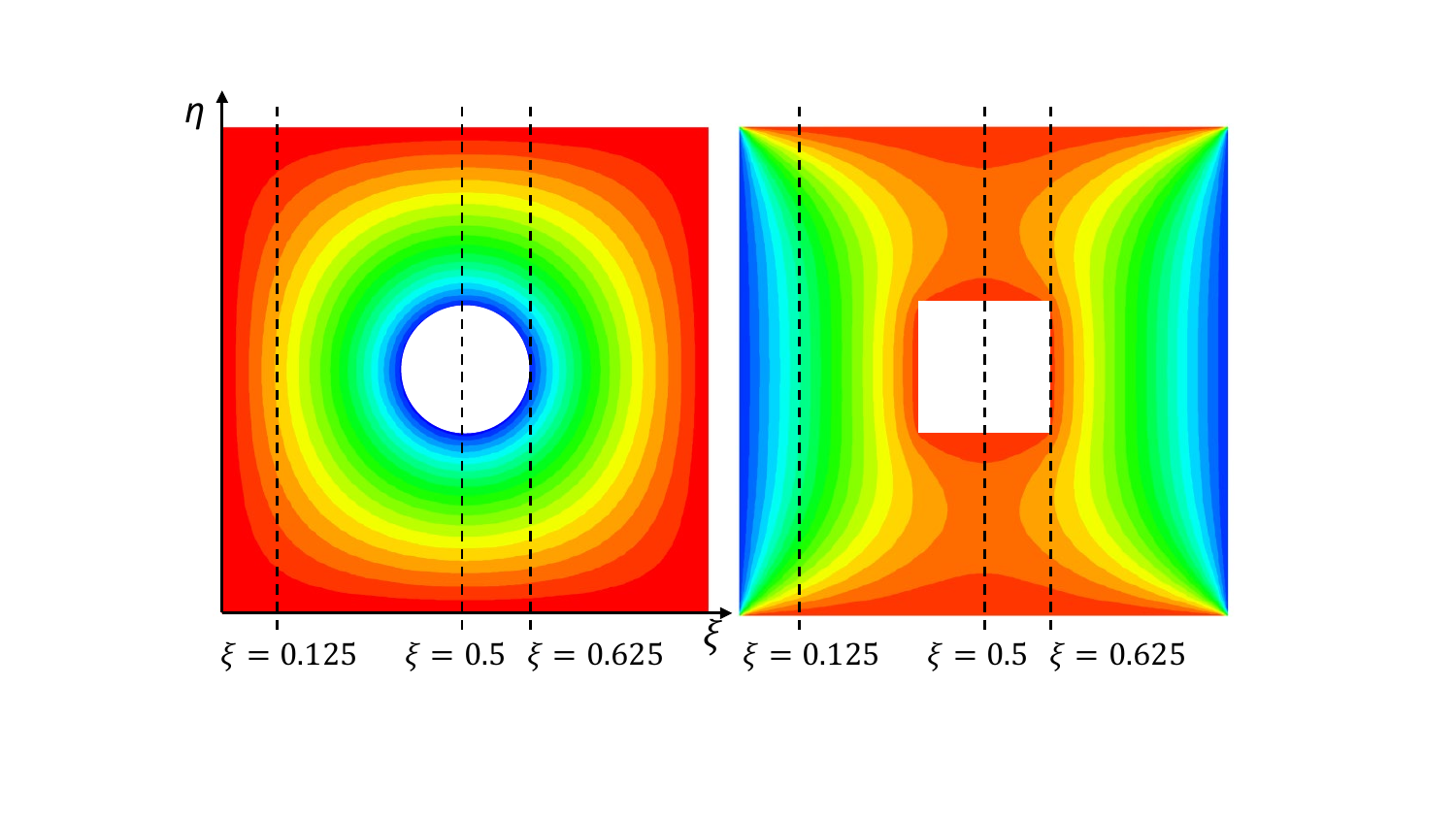}
	\caption{{Locations for temperature extraction.}}
	\label{figs:plate locations}
\end{figure}

Figure~\ref{figs:squareplates} compares the temperature distributions of the ground truth, DDPM, and LDM for plates with square holes. Groups (a), (b), and (c) correspond to plates with square holes located at different positions. Both the DDPM and LDM predictions show good overall agreement with the ground truth.  However, the LDM yields a more accurate temperature distribution across most of the domain compared to the DDPM, with its accuracy slightly decreasing near edges and corners.   This observation is further supported in Figure~\ref{figs:squareplateserror}, where the prediction errors—highlighted in red—are predominantly concentrated in the computational regions adjacent to edges and corners.  As discussed above,  this discrepancy can be partly attributed to certain losses at the boundary during the compression of the data into the latent space and its subsequent reconstruction.  For a more detailed assessment of the temperature predictions, as shown in Figure~\ref{figs:plate locations}, temperature profiles are extracted at three locations: $\xi = 0.125$, 0.5, and 0.625, enabling a direct comparison.

\begin{figure}[htb!]
	\centering
	\includegraphics[trim = 2cm 0cm 1cm 0cm, clip,width=\textwidth]{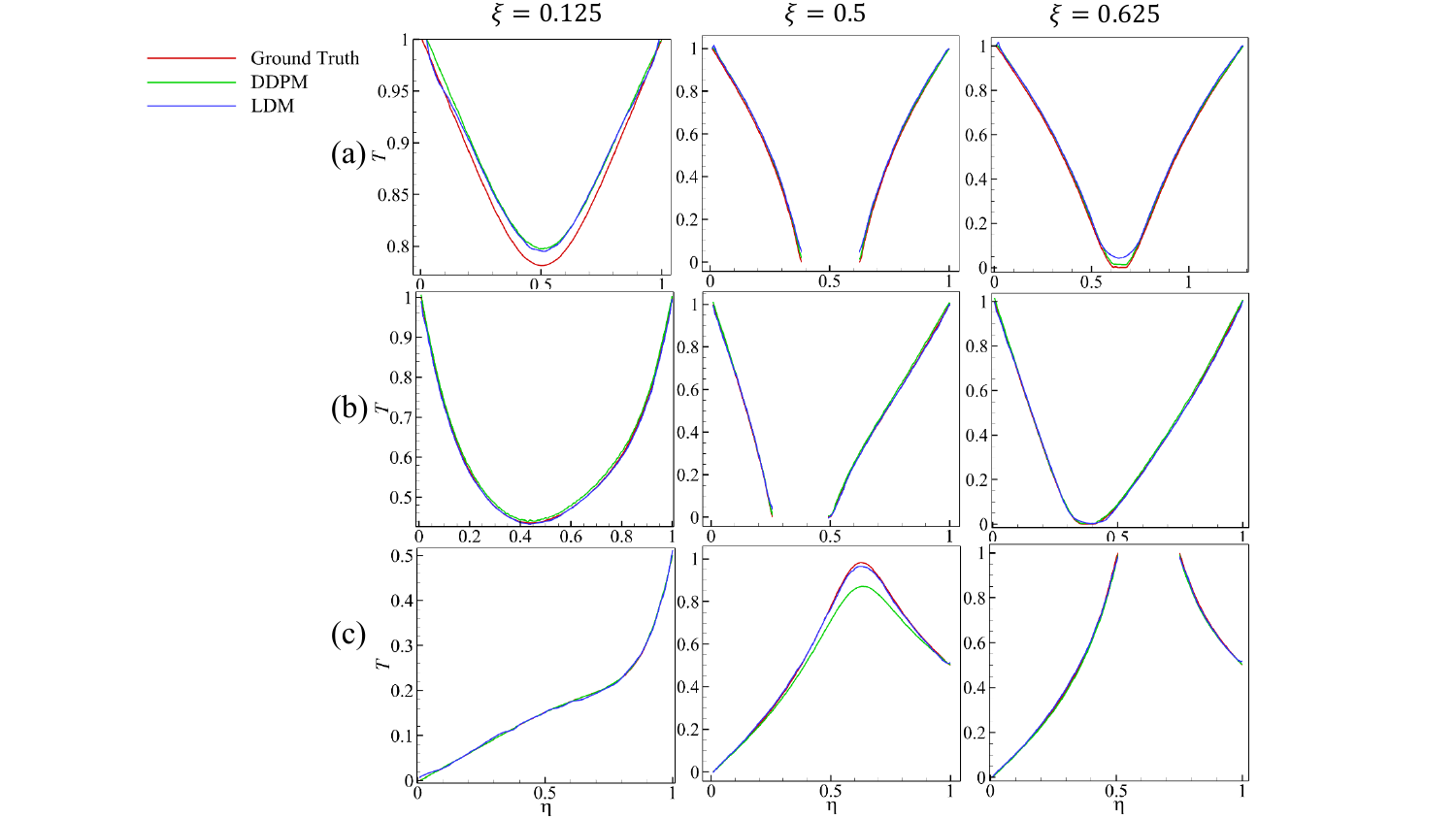}
	\caption{{Results of plate (with circular holes) temperature distributions at locations $\xi$ = 0.125, $\xi$ = 0.5, and $\xi$ = 0.625.}}
	\label{figs:circleplateT}
\end{figure}
\begin{figure}[htb!]
	\centering
	\includegraphics[trim = 2cm 0.1cm 1cm 0.2cm, clip,width=\textwidth]{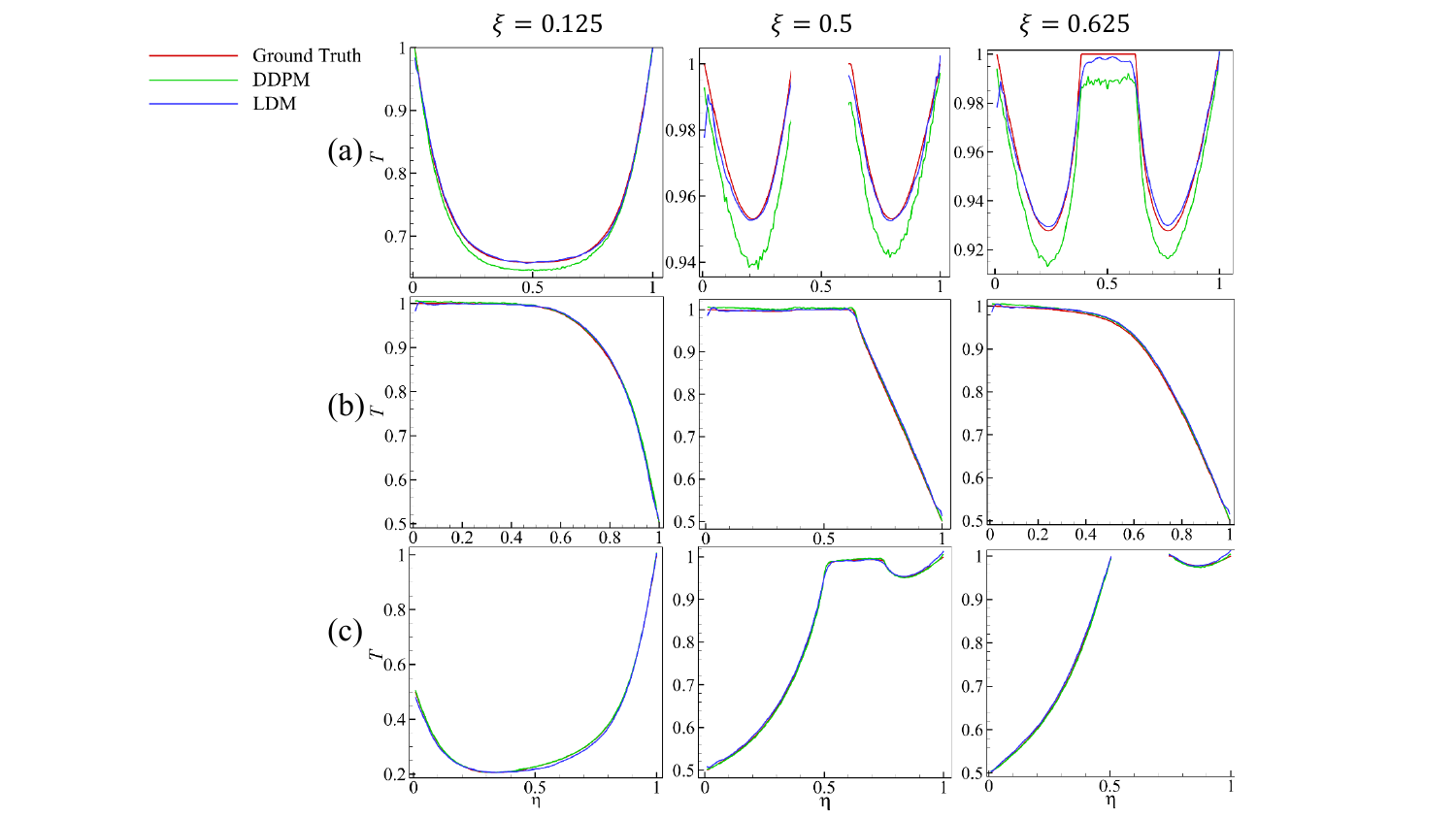}
	\caption{{Results of plate (with square holes) temperature distributions at locations $\xi$ = 0.125, $\xi$ = 0.5, and $\xi$ = 0.625.}}
	\label{figs:squareplateT}
\end{figure}

Figures~\ref{figs:circleplateT}-~\ref{figs:squareplateT} compare the temperature distributions along the normal direction at these three positions. As shown in Figure~\ref{figs:circleplateT}(a), the prediction demonstrates broad consistency with the ground truth, with the exception of a localized discrepancy at a mid-point along the $\eta$ direction. In Figures~\ref{figs:circleplateT}(b) and ~\ref{figs:circleplateT}(c), both DDPM and LDM exhibit good agreement with the ground truth overall, except at $\xi = 0.5$ in Figure~\ref{figs:circleplateT}(c), where a noticeable discrepancy appears between the DDPM results and the ground truth. Meanwhile, small fluctuations are visible near the inner hole in the LDM results near $\xi = 0.5$ in Figures~\ref{figs:circleplateT}(a) and (b), which may stem from errors introduced during the compression and reconstruction stages.

For the plate with a square hole in Figure~\ref{figs:squareplateT}(a) at $\xi = 0.125$, the predictions from both DDPM and LDM agree well with the ground truth, with LDM showing particularly high accuracy. At locations $\xi = 0.5$ and $\xi = 0.625$, LDM generally outperforms DDPM. However, in the vicinity of the square hole, both models show room for improvement, which could be addressed by explicitly imposing inner‑hole boundary conditions.
Moreover, the DDPM predictions exhibit noticeable fluctuations along the $\eta$ direction, particularly near the inner hole. This is attributed to the model’s operation in the high‑dimensional pixel space, where the inherent stochasticity of the diffusion sampling process and the difficulty in stabilizing high‑frequency details are directly manifested in the output field. As a result, local oscillations become visually prominent, especially near geometric discontinuities. Overall, both DDPM and LDM achieve a high accuracy compared with the ground truth in Figures~\ref{figs:squareplateT}(b) and (c). The only notable exception is a minor disturbance in the LDM results near the edge of the computational domain, which can again be linked to the compression and reconstruction processes.
\subsection{Predictions of airfoil flow fields}
\begin{figure}[htb!]
	\centering
	\includegraphics[trim = 2cm 0.2cm 2cm 0cm, clip,width=\textwidth]{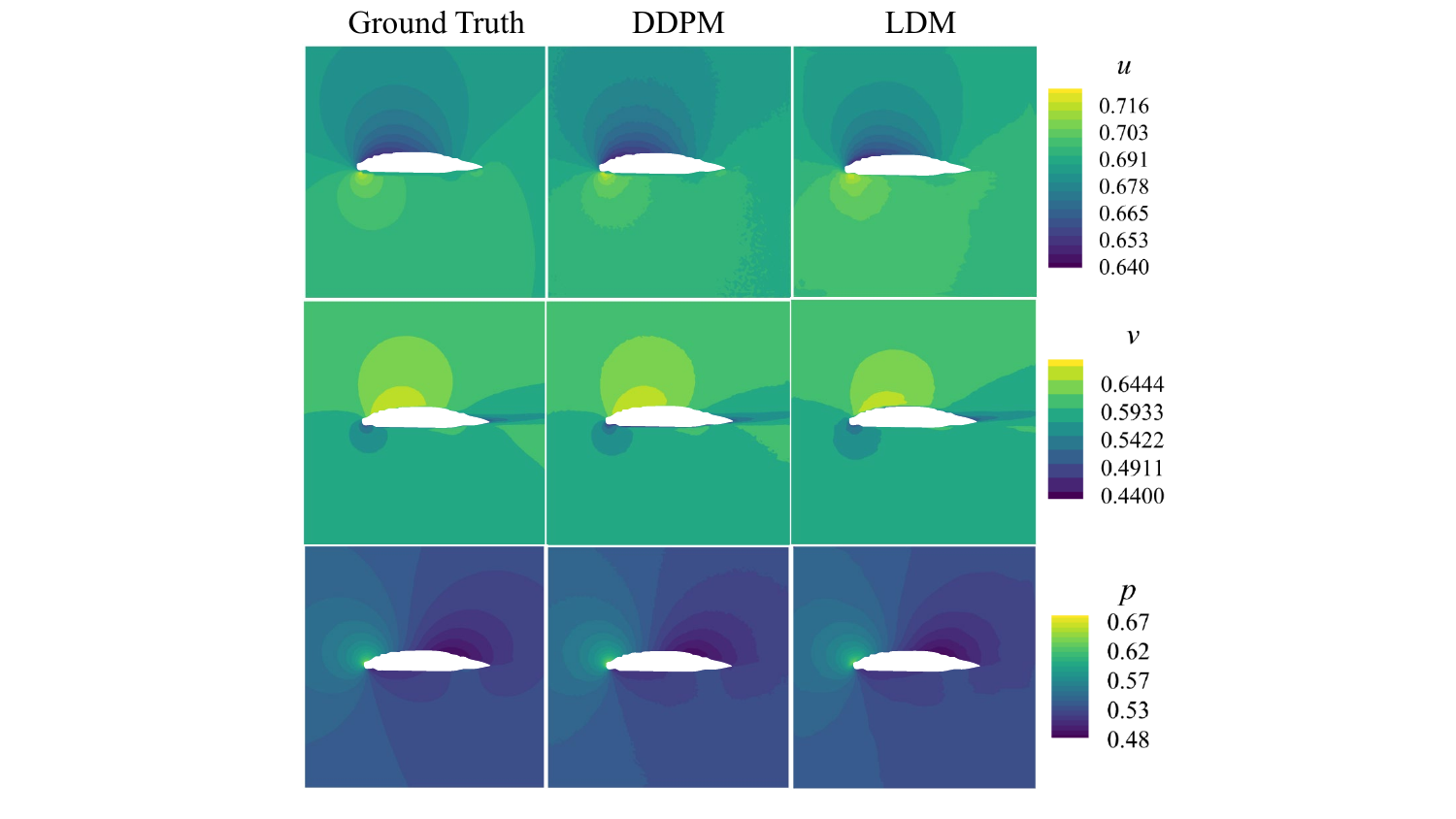}
	\caption{{Comparison of the flow fields between the ground truth, DDPM, and LDM around the AH94156 airfoil at $\alpha'$ = 8.27°.}}
	\label{figs:airfoiluvp}
\end{figure}

\begin{table}
\centering
\caption{The average errors of  DDPM and LDM for the airfoil flow fields of all test cases.}
\resizebox{0.6\textwidth}{!}{%
\tiny 
\begin{tabular}{ccccc}
\hline
Models   & $\bm{\zeta}$     & $\bm{\zeta}_{u}$  & $\bm{\zeta}_{v}$ & $\bm{\zeta}_{p}$         \\
\hline
 DDPM & 0.014802 & 0.021460 & {\color{blue}0.011183}& 0.011764\\
 LDM & {\color{blue}0.011624} & {\color{blue}0.008728} & 0.021445
& {\color{blue}0.004700}\\

\hline
\end{tabular}%
}
\label{tab:airfoil error}
\end{table}
The models are then evaluated on airfoil flows and Table~\ref{tab:airfoil error} illustrates the prediction errors by DDPM and LDM across all test cases, with the lower errors marked in blue. LDM achieves a slightly lower global prediction error than DDPM (0.011624 vs. 0.014802). It demonstrates superior accuracy in predicting the $u$ velocity and pressure fields, whereas DDPM proves more accurate for the $v$ velocity component. Figure~\ref{figs:airfoiluvp} compares the velocity and pressure flow fields around the AH94156 airfoil at $\alpha'$ = 8.27°, contrasting the ground truth with the predictions of DDPM and LDM. The first column shows the ground truth by CFD, with the next two columns displaying the predicted flow fields by DDPM and LDM.

\begin{figure}[htb!]
	\centering
	\includegraphics[trim = 2cm 2cm 2cm 1.3cm, clip,width=\textwidth]{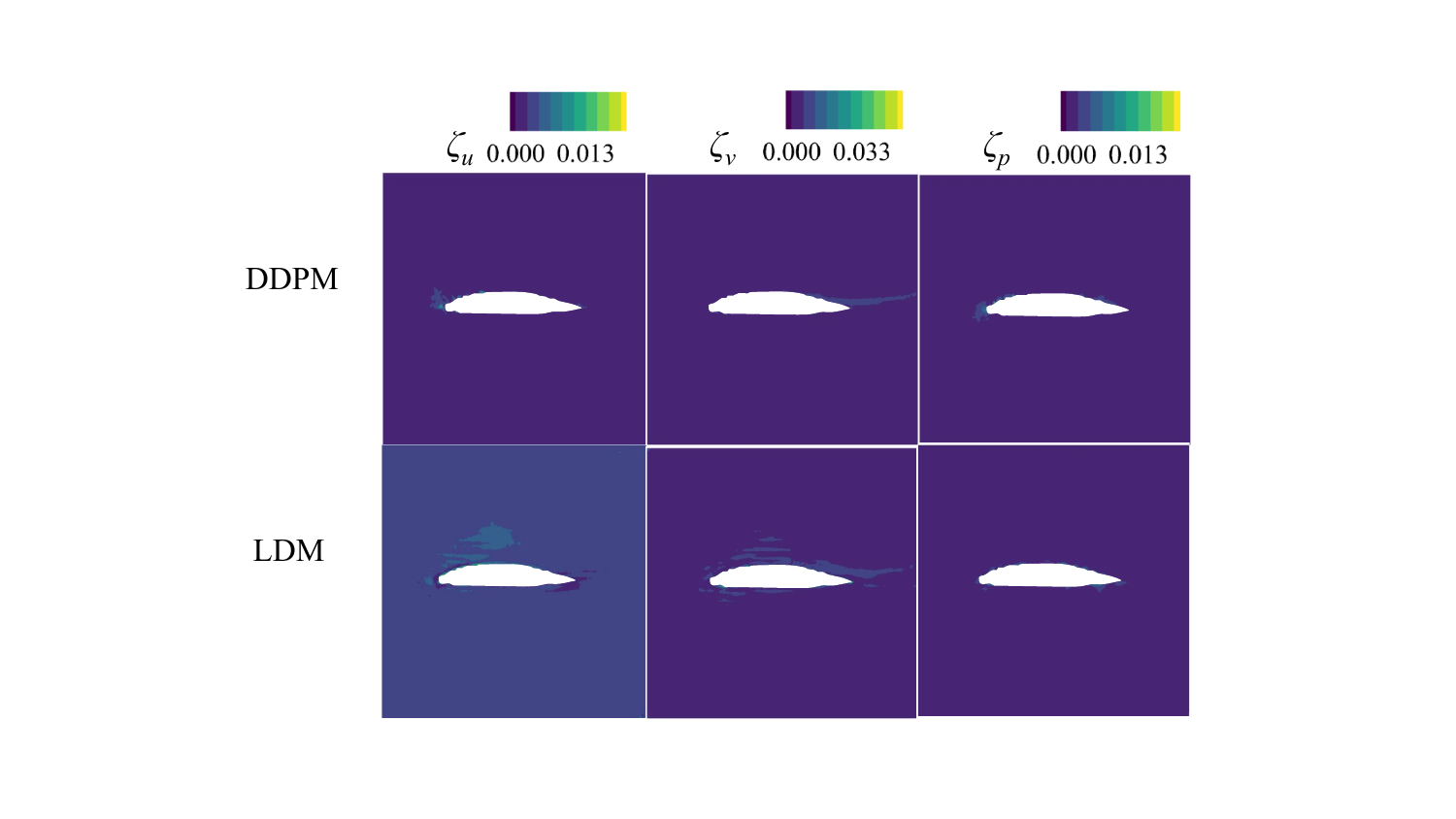}
	\caption{{Comparison of the prediction errors between  DDPM and LDM around the AH94156 airfoil at $\alpha'$ = 8.27°.}}
	\label{figs:airfoilerror}
\end{figure}

As shown in the $u$ velocity fields in Figure~\ref{figs:airfoiluvp}, the DDPM achieves a closer match to the ground truth, successfully reconstructing fine-scale flow structures—especially in regions with sharp gradients and within the wake. In comparison, the LDM exhibits a slight degradation in prediction accuracy. Nevertheless, it retains overall consistency with the ground truth in terms of both position and shape of key flow features, indicating that the LDM remains a viable and effective approach for flow field prediction. Meanwhile, although the DDPM prediction yields marginally closer numerical agreement with the ground truth, its contour boundaries exhibit noticeable graininess, suggesting a trade-off between prediction accuracy and visual smoothness. The graininess observed in the DDPM profile originates from the random denoising process within the diffusion model. Minor errors in noise prediction at each sampling step accumulate progressively, ultimately leading to high-frequency fluctuations in the output field. While this phenomenon can enhance fine details in certain scenarios, it introduces undesirable noise at phase boundaries in this case. For the $v$ velocity fields, both models capture the overall structure of the flow around the airfoil and the prediction results demonstrate an enhanced accuracy in the high $v$ velocity fields above the airfoil surface by DDPM. For the pressure field $p$, both DDPM and LDM reconstruct the complex pressure distribution with high efficiency. The predictions exhibit high fidelity to the ground truth with minimal error, underscoring the strong potential of diffusion models for high fidelity reconstruction and generation of fluid dynamic data. As shown in Figure~\ref{figs:airfoilerror} about the prediction errors of DDPM and LDM, the errors mainly lie in the fields around the airfoil and DDPM has a lower error, especially in the velocity field $u$. From the prediction results in Figure~\ref{figs:airfoiluvp} we know that the prediction errors of LDM are within an acceptable range and therefore another case by LDM is picked to compare with the ground truth.

\begin{figure}[htb!]
	\centering
	\includegraphics[trim = 1cm 0.5cm 1cm 0.3cm, clip,width=\textwidth]{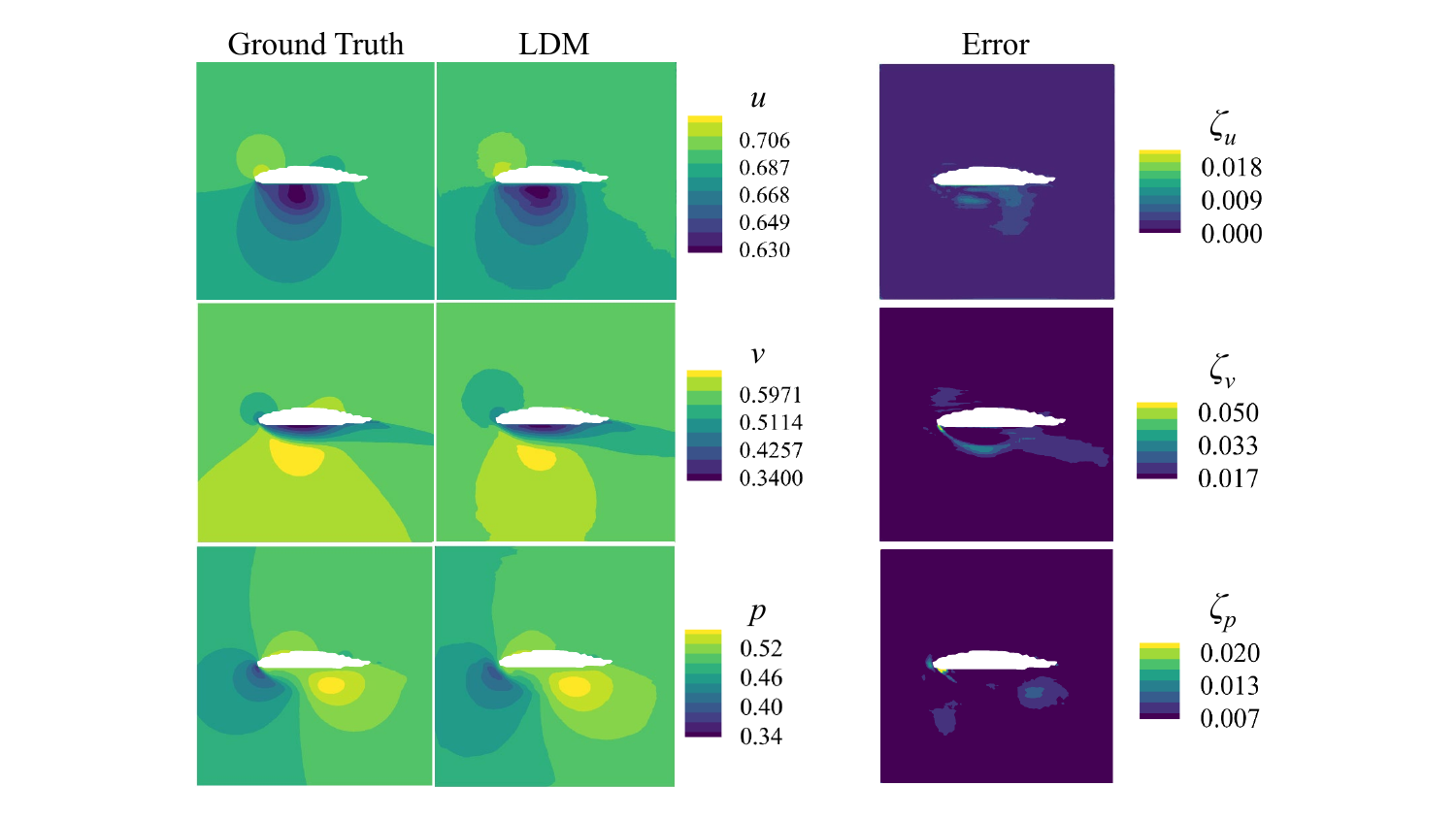}
	\caption{{Comparison of the prediction results between the ground truth and LDM around the AH63k127 airfoil at $\alpha'$ = -13.59°.}}
	\label{figs:airfoiluvperror}
\end{figure}
Figure~\ref{figs:airfoiluvperror} presents the flow fields around the AH63k127 airfoil at $\alpha'$ = -13.59°, comparing the ground truth with the LDM results. The results indicate that the LDM can achieve outcomes relatively close to the ground truth. However, some deviations are observed in the $v$ velocity, and the predicted flow exhibits slight distortions, which are likely introduced during the compression and reconstruction stages. Additionally, prediction errors are primarily concentrated near the airfoil surface.

\subsection{Predictions of hypersonic flow fields}
\begin{figure}[htb!]
	\centering
	\includegraphics[trim = 2cm 3.7cm 2cm 2.1cm, clip,width=\textwidth]{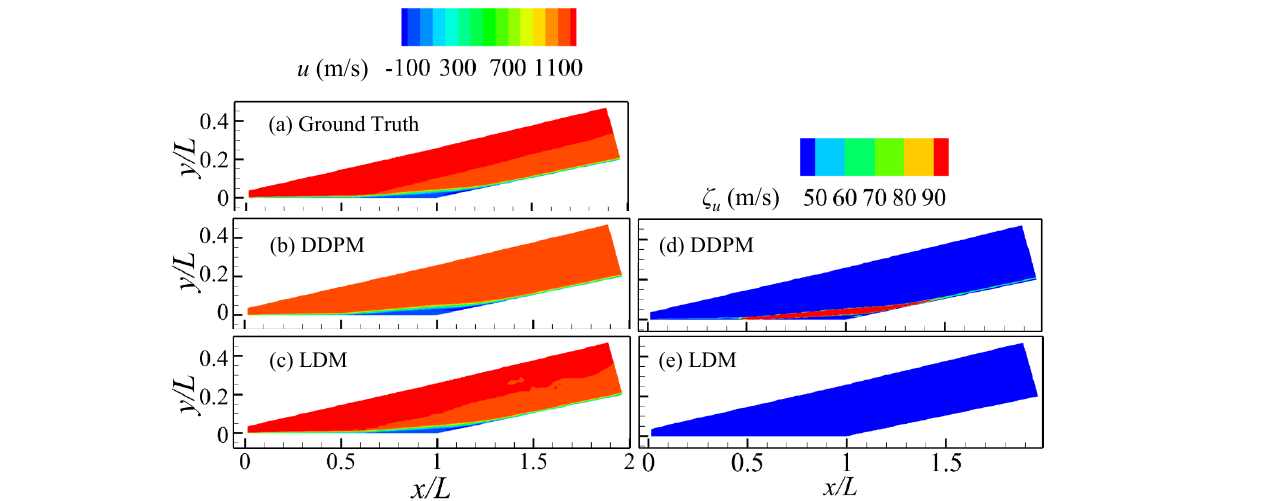}
	\caption{{Comparison of $u$ velocity fields of ground truth, DDPM and LDM at $Ma=5.46$, $Re=4.79\times10^6$, $\alpha=11.7^\circ$, $ T_{\infty}$ = 124 K, and  $ T_w /T_0 $ = 0.56.  (a) Ground truth; (b) DDPM; (c) LDM; (d) and (e) corresponding prediction errors of DDPM and LDM, respectively.}}
	\label{figs:srampU}
\end{figure}
\begin{figure}[htb!]
	\centering
	\includegraphics[trim = 2cm 4cm 2cm 2cm, clip,width=\textwidth]{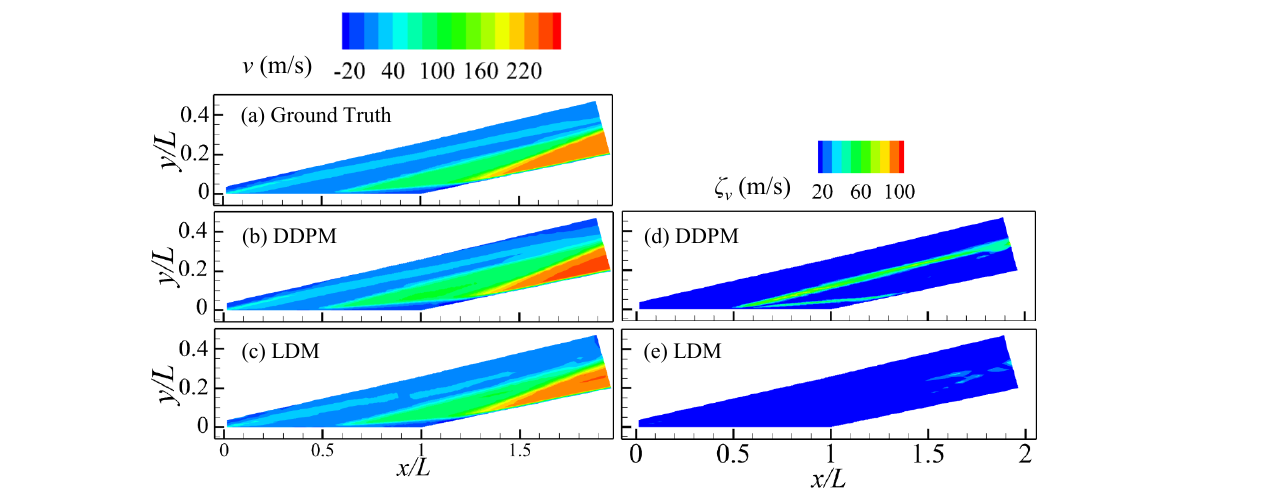}
	\caption{{Comparison of $v$ velocity fields of ground truth, DDPM and LDM at $Ma=5.46$, $Re=4.79\times10^6$, $\alpha=11.7^\circ$, $ T_{\infty}$ = 124 K, and  $ T_w /T_0 $ = 0.56.  (a) Ground truth; (b) DDPM; (c) LDM; (d) and (e) corresponding prediction errors of DDPM and LDM, respectively.}}
	\label{figs:srampV}
\end{figure}
\begin{figure}[htb!]
	\centering
	\includegraphics[trim = 2cm 4.4cm 2cm 1.4cm, clip,width=\textwidth]{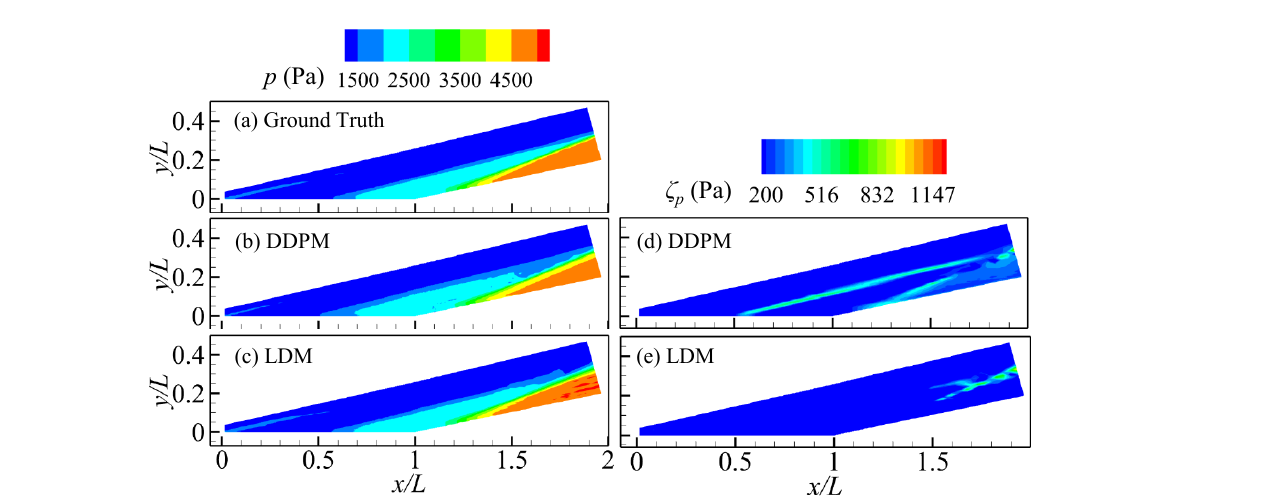}
	\caption{{Comparison of $p$ prssure fields of ground truth, DDPM and LDM at $Ma=5.46$, $Re=4.79\times10^6$, $\alpha=11.7^\circ$, $ T_{\infty}$ = 124 K, and  $ T_w /T_0 $ = 0.56.  (a) Ground truth; (b) DDPM; (c) LDM; (d) and (e) corresponding prediction errors of DDPM and LDM, respectively.}}
	\label{figs:srampP}
\end{figure}
\begin{figure}[htb!]
	\centering
	\includegraphics[trim = 2cm 4.5cm 2cm 2.1cm, clip,width=\textwidth]{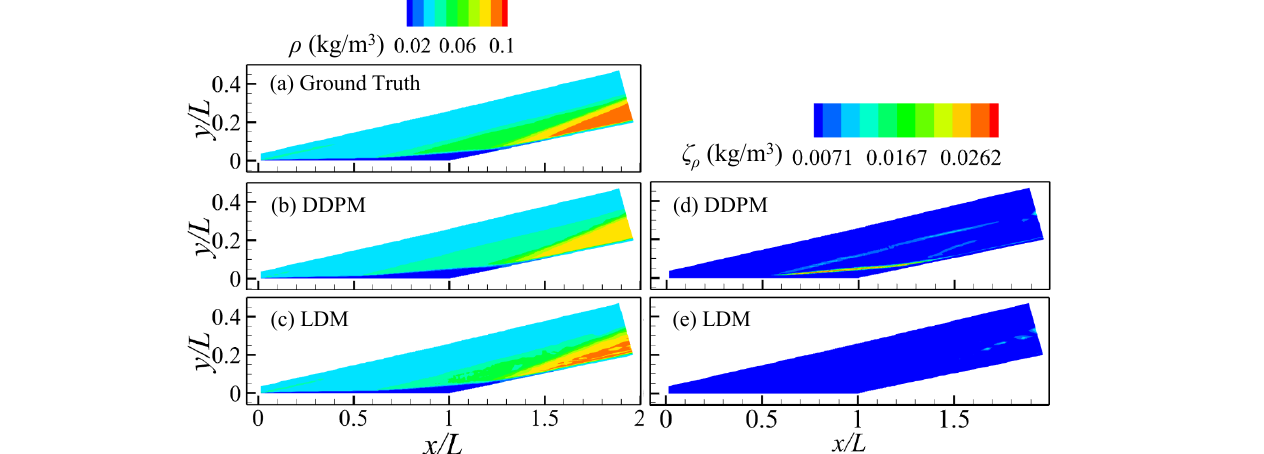}
	\caption{{Comparison of $\rho$ density fields of ground truth, DDPM and LDM at $Ma=5.46$, $Re=4.79\times10^6$, $\alpha=11.7^\circ$, $ T_{\infty}$ = 124 K, and  $ T_w /T_0 $ = 0.56.  (a) Ground truth; (b) DDPM; (c) LDM; (d) and (e) corresponding prediction errors of DDPM and LDM, respectively.}}
	\label{figs:srampROU}
\end{figure}

Figures~\ref{figs:srampU}-~\ref{figs:srampROU} present a comparative analysis of the velocity, pressure, and density flow fields between the ground truth and the predictions from the DDPM and LDM, under the simulation conditions of $Ma=5.46$, $Re=4.79\times10^6$, $\alpha=11.7^\circ$, $T_{\infty}=124$ K, and $T_w/T_0=0.56$. 

As illustrated in Figure~\ref{figs:srampU} for the streamwise velocity $u$, the LDM prediction (Fig.~\ref{figs:srampU}(c)) exhibits closer agreement with the ground truth than the DDPM result (Fig.~\ref{figs:srampU}(b)), particularly downstream of the reattachment shock and separation shock. The error distributions in Figures~\ref{figs:srampU}(d) and ~\ref{figs:srampU}(e) further reveal that the primary discrepancies for DDPM are concentrated near the separated shear layer, whereas the LDM yields significantly lower errors.

Similar error characteristics are observed in the normal velocity $v$ in Figure~\ref{figs:srampV}. The errors by DDPM (Fig.~\ref{figs:srampV}(d)) are prominent near the separated shear layer, separation shock, downstream of the reattachment shock, and along the slip line (see Figure~\ref{figs:ramp1}). In contrast, the LDM prediction (Fig.~\ref{figs:srampV}(e)) shows only minor errors near the slip line. A direct comparison between Figures~\ref{figs:srampV}(a)-~\ref{figs:srampV}(c) confirms that the LDM captures the flow features more accurately, especially in the region influenced by the reflected expansion wave.

From the pressure fields in Figures~\ref{figs:srampP}(a)-~\ref{figs:srampP}(c), LDM prediction captures the structure of the reflected expansion wave and the area between the wall surface and the separation shock more faithfully.  The error analysis indicates that DDPM discrepancies are located near the separation shock, reattachment shock, downstream of the reattachment shock, and along the slip line, while LDM errors are largely confined to the slip line. 

Finally, concerning the density fields in Figure~\ref{figs:srampROU}, the errors of LDM mainly lie in the area near the slip line, while the primary errors for DDPM remain concentrated within the separated shear layer, separation shock, and reattachment shock.

\begin{figure}[htb!]
	\centering
	\includegraphics[trim = 5cm 1.5cm 6cm 12.5cm, clip,width=0.75\textwidth]{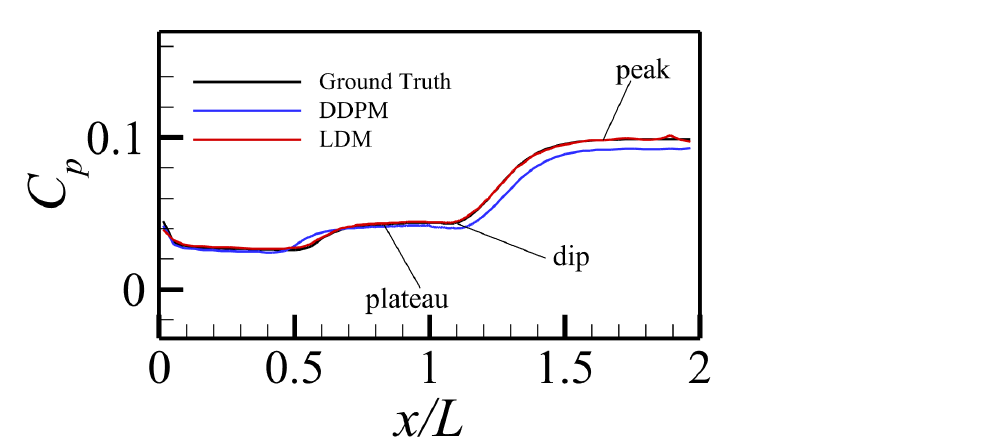}
	\caption{Comparison of wall pressure of the ground truth, DDPM and LDM at $Ma=5.46$, $Re=4.79\times10^6$, $\alpha=11.7^\circ$, $ T_{\infty}$ = 124 K, and  $ T_w /T_0 $ = 0.56.}
	\label{figs:cpdownsample}
\end{figure}
Figure~\ref{figs:cpdownsample} compares the surface pressure coefficient $C_{p}$ distributions from the ground truth, DDPM predictions, and LDM predictions. Overall, both models align closely with the ground truth, but the LDM prediction more accurately captures key surface flow features—notably the local pressure dip near the corner. This dip corresponds to the locally intensified pressure gradient, which strengthens with increasing ramp angle until the reverse‑flow boundary layer can no longer sustain it, ultimately leading to secondary separation \cite{hao2021occurrence}.
\begin{figure}[htb!]
	\centering
	\includegraphics[trim = 0.2cm 1.5cm 0.5cm 1.3cm, clip,width=\textwidth]{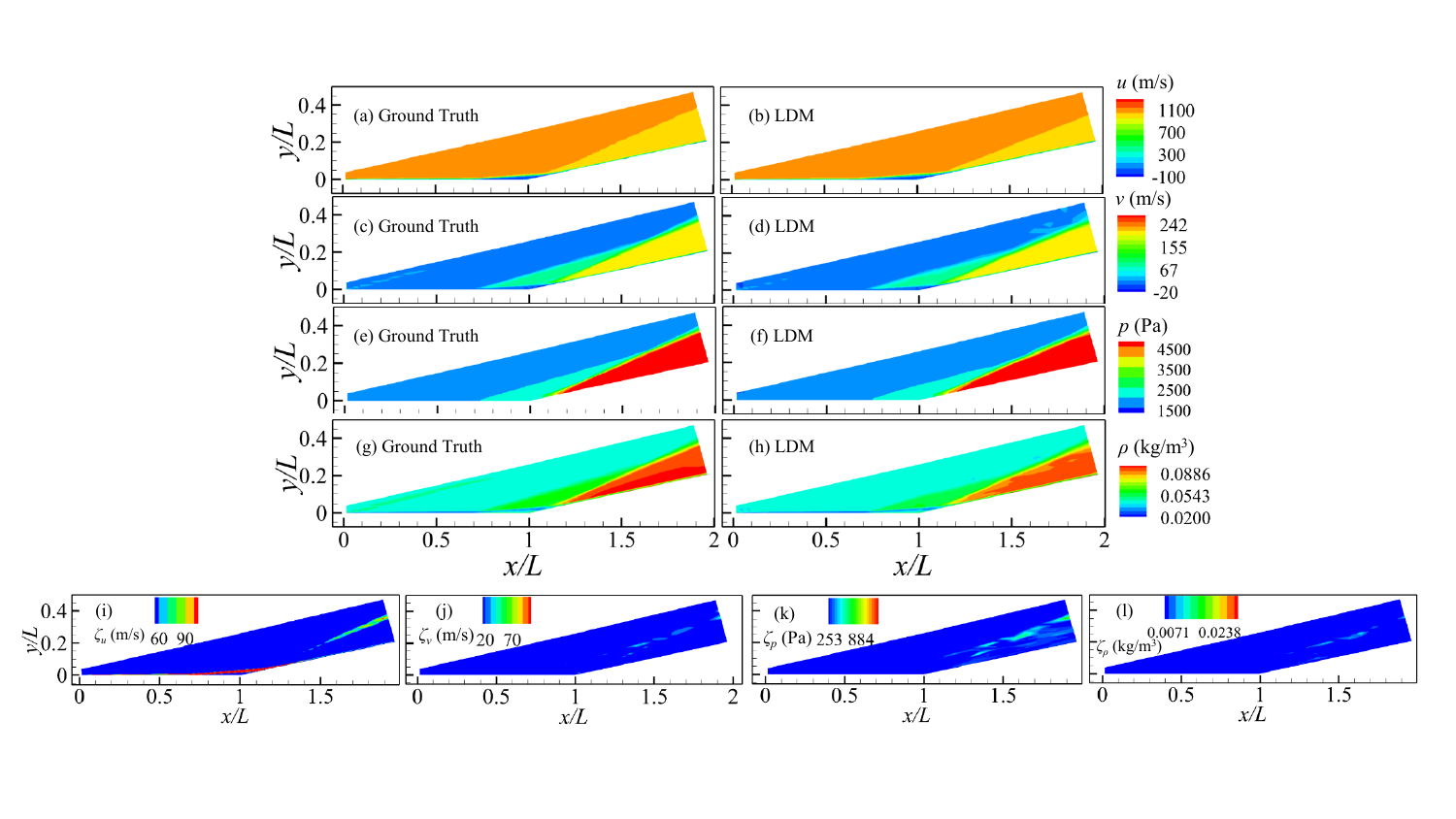}
	\caption{{Comparison of flow fields of the ground truth and LDM under $ Ma $ = 4.39,  $ Re   = 4.45\times 10^6$,  $ \alpha  $ = 12.1°, $ T_{\infty}$ = 135 K and  $ T_w /T_0 $ = 0.24.  (a),(c),(e),(g): Ground truth; (b),(d),(f),(h): LDM; (i)-(l): corresponding prediction errors of LDM.}}
	\label{figs:smallrampfield}
\end{figure}

\begin{figure}[htb!]
	\centering
	\includegraphics[trim = 5cm 7.5cm 5cm 8.5cm, clip,width=0.8\textwidth]{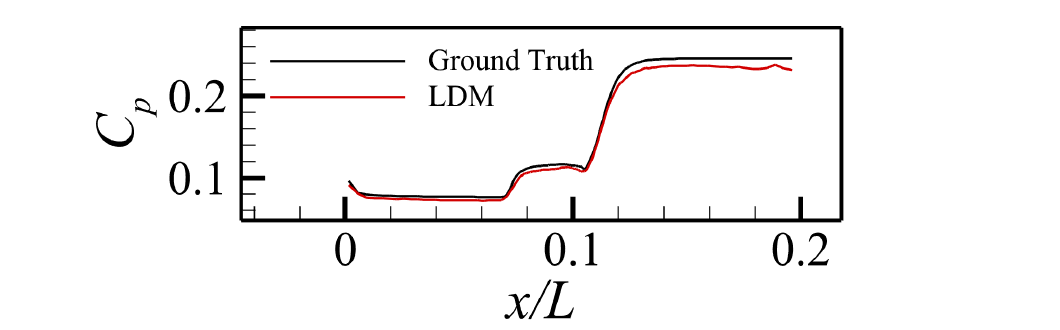}
	\caption{Comparison of wall pressure of the ground truth and LDM with $ Ma $ = 4.39,  $ Re   = 4.45\times 10^6$,  $ \alpha  $ = 12.1°, $ T_{\infty}$ = 135 K and  $ T_w /T_0 $ = 0.24. }
	\label{figs:CPSMALLRAMP}
\end{figure}
To further analyze the prediction performance of LDM, Figure~\ref{figs:smallrampfield} compares another flow fields obtained from the ground truth and the LDM under the conditions of $ Ma $ = 4.39,  $ Re   = 4.45\times 10^6$,  $ \alpha  $ = 12.1°, $ T_{\infty}$ = 135 K and  $ T_w /T_0 $ = 0.24. The LDM results exhibit good agreement with the ground truth, except for certain prediction discrepancies observed in the streamwise velocity near the separated shear layer, as illustrated in Figure~\ref{figs:smallrampfield}(i).  Figure~\ref{figs:CPSMALLRAMP} compares the surface pressure coefficient between the ground truth and the LDM results. In this flow configuration, the dip of the pressure coefficient near the compression corner is more pronounced.
\subsection{Predictions of full hypersonic flow fields by LDM}
Based on the above sections, both DDPM and LDM have been applied to thermal and incompressible as well as hypersonic flow fields, with downsampling employed in the hypersonic cases to save training sources. These applications demonstrate the generalization of the LDM. Accordingly, in this section, the LDM is selected to perform predictions for full-resolution hypersonic flow fields.
\begin{figure}[htb!]
	\centering
	\includegraphics[trim = 0.2cm 0.7cm 0.5cm 0.3cm, clip,width=\textwidth]{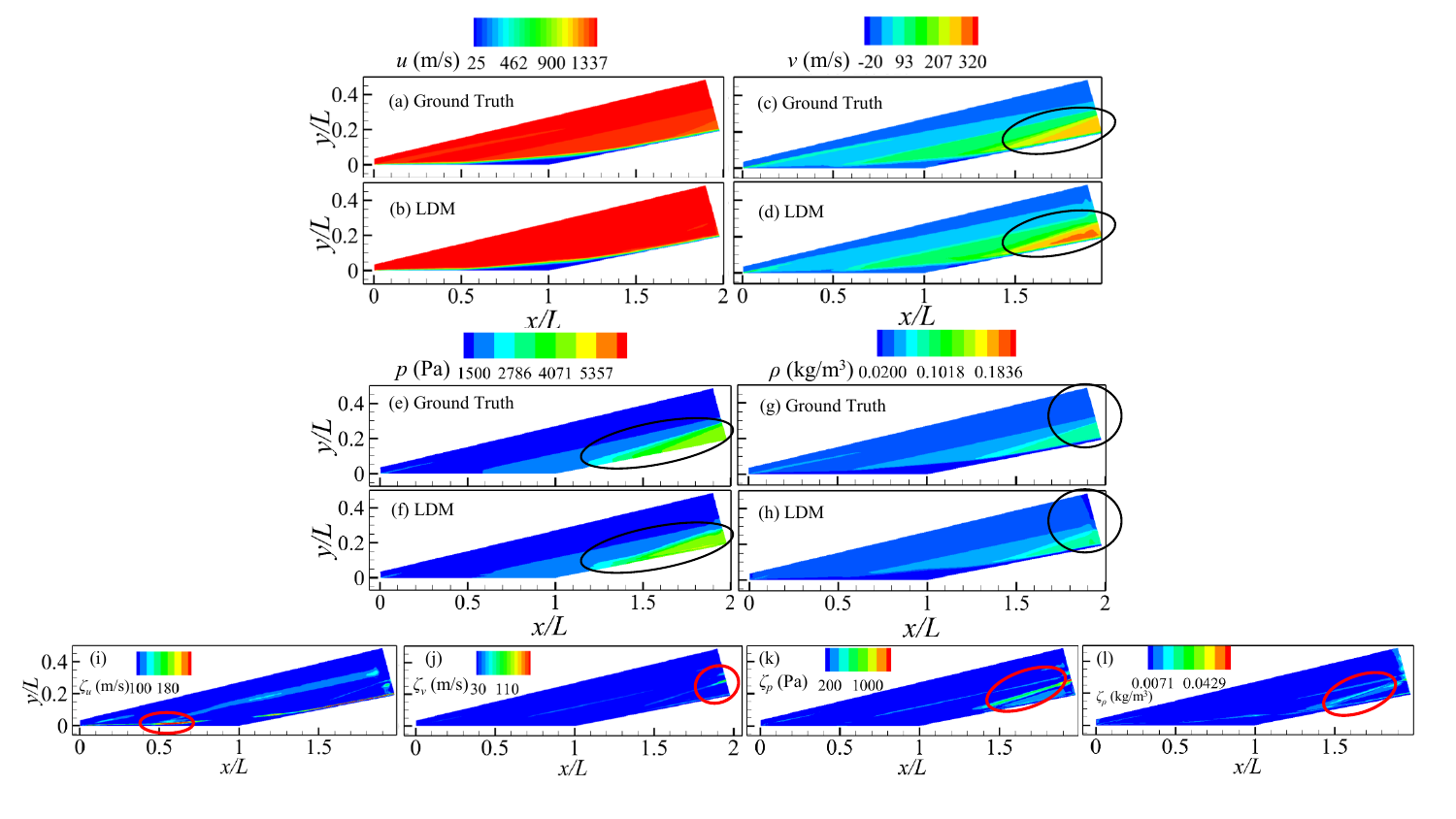}
	\caption{{Comparison of flow fields of the ground truth and LDM under $ Ma $ = 6.61,  $ Re   = 4.92\times 10^6$,  $ \alpha  $ = 11.1°, $ T_{\infty}$ = 113 K and  $ T_w /T_0 $ = 0.80.  (a),(c),(e),(g): Ground truth; (b),(d),(f),(h): LDM; (i)-(l): corresponding prediction errors of LDM.}}
	\label{figs:LDMFULLRAMP}
\end{figure}

Figure~\ref{figs:LDMFULLRAMP} presents a comparison between the ground truth and the flow fields predicted by LDM.  Overall, the LDM predictions align well with the ground truth, except for some localized discrepancies highlighted by circles. As shown in Figures~\ref{figs:LDMFULLRAMP}(c)–~\ref{figs:LDMFULLRAMP}(h), differences in the normal velocity $v$, pressure and density fields primarily appear downstream, particularly in regions associated with the slip line and the reflected expansion wave. This observation can be further confirmed in Figures~\ref{figs:LDMFULLRAMP}(j)-~\ref{figs:LDMFULLRAMP}(l), where the prediction errors are predominantly concentrated within these areas located near the boundary. These discrepancies may be attributed to the following factors. First, as a data-driven model, the LDM lacks the capability to enforce physical boundary conditions as rigorously as conventional numerical solvers, which can result in slight deviations along boundaries. Second, the denoising process does not explicitly account for the physical relationships between upstream and downstream flow regions, rendering downstream areas more susceptible to generation uncertainties. Furthermore, information loss during the autoencoder’s compression and reconstruction stages may amplify the difficulty of recovering fine flow details in these zones.

A key observation from Figure~\ref{figs:LDMFULLRAMP}(i) is that the prediction errors for the $u$ velocity field are predominantly localized at the leading edge of the separated shear layer. This distribution contrasts with previous studies~\cite{jia2025prediction}, where errors typically concentrate around shock waves and high-pressure gradient regions. This discrepancy highlights a fundamental difference in model behavior. Traditional regression models such as Convolutional Neural Networks (CNNs) predict flow fields through point-by-point mapping. When solving the discontinuous features like shock waves, they tend to output statistically averaged physical quantities, resulting in smeared shock profiles and localized errors—a limitation that often necessitates physics-informed loss functions for mitigation. In contrast, the LDM learns the underlying probability distribution of the flow field and generates results through an iterative denoising process initiated from random noise. In this process, early denoising steps capture large-scale structures such as shock waves, while later steps refine and sharpen these features. Consequently, the LDM produces physically sharper and more realistic shock waves, leading to significantly reduced errors in such critical regions. Therefore, while the LDM exhibits slightly higher global errors than the vision transformer(predictions for velocity and pressure fields)~\cite{jia2025prediction}, as quantified in Table~\ref{tab:fullramp error}, it achieves this by prioritizing fidelity in critical regions.


%
\begin{table}
\centering
\caption{The average errors of the vision transformer and LDM for the compression ramp flow fields of all test cases.}
\resizebox{0.76\textwidth}{!}{%
\tiny 
\begin{tabular}{cccccc}
\hline
Models   & $\bm{\zeta}$     & $\bm{\zeta}_{u}$  & $\bm{\zeta}_{v}$ & $\bm{\zeta}_{p}$ & $\bm{\zeta}_{\rho}$        \\
\hline
 Vision transformer\cite{jia2025prediction}& 0.015212 & 0.024748 & 0.012597 & 0.008290& --\\
 LDM & 0.020008& 0.034432 & 0.016308 & 0.018114& 0.011180
\\

\hline
\end{tabular}%
}
\label{tab:fullramp error}
\end{table}

\begin{figure}
	\centering
         \begin{subfigure}[b]{1\textwidth}
		\centering
		\includegraphics[trim = 2.8cm 1.7cm 5cm 13cm, clip, width=0.8\textwidth]{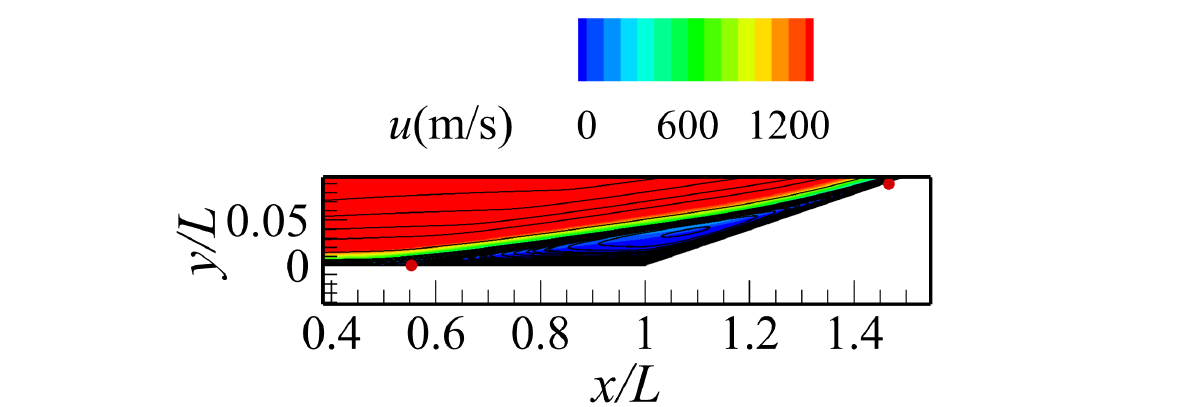}
        \caption{Ground truth}
        \end{subfigure}
                 \begin{subfigure}[b]{1\textwidth}
		\centering
		\includegraphics[trim = 2.8cm 1.7cm 5cm 15cm, clip, width=0.8\textwidth]{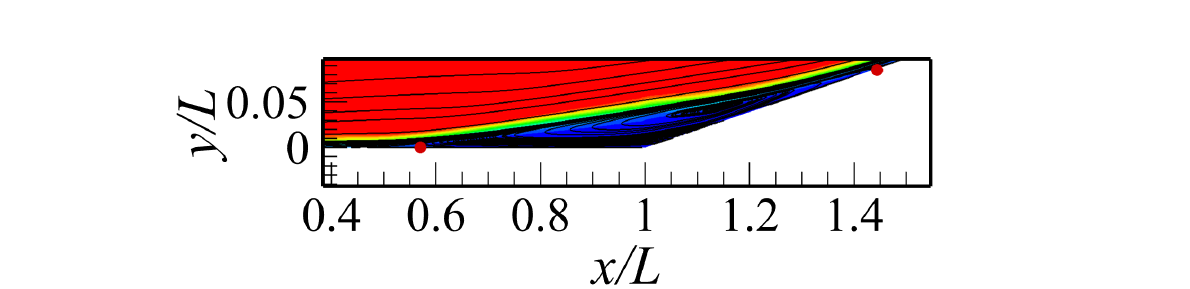}
        \caption{LDM}
        \end{subfigure}
	\caption{Comparison of flow field with streamlines of the ground truth and LDM under $ Ma $ = 6.61,  $ Re   = 4.92\times 10^6$,  $ \alpha  $ = 11.1°, $ T_{\infty}$ = 113 K and  $ T_w /T_0 $ = 0.80.  (a) Ground truth; (b) LDM. Closed circles: separation and reattachment points.}
    \label{figs:liuchang}
\end{figure}

Figure~\ref{figs:liuchang} compares the separation regions with streamlines between the ground truth and the LDM prediction, showing good agreement in the vortex core location. Table~\ref{tab:separation regions} further compares the separation and reattachment locations. It can be observed that the LDM achieves a separation length deviation of only 4.28\%, outperforming the vision transformer model~\cite{jia2025prediction}, which records a slightly higher deviation of 4.91\%. This improvement suggests that the LDM is more effective in preserving critical flow features.  
\begin{table}
\centering
\caption{Analysis of separation regions.}
\resizebox{0.85\textwidth}{!}{%
\tiny 
\begin{tabular}{ccccc}
\hline
Method   & $x/L$, separation point             &$x/L$, reattachment point       &separation length \\
\hline
Ground truth & 0.55272 & 1.46610 & 0.91338  \\
Vision transformer\cite{jia2025prediction} & 0.57127 & 1.43978 & 0.86851(4.91\%deviation)  \\
LDM & 0.56975 & 1.44408 & 0.87433(4.28\%deviation)
   \\

\hline
\end{tabular}%
}
\label{tab:separation regions}
\end{table}

\begin{figure}[htb!]
	\centering
	\includegraphics[trim = 5cm 1cm 5cm 15.5cm, clip,width=0.8\textwidth]{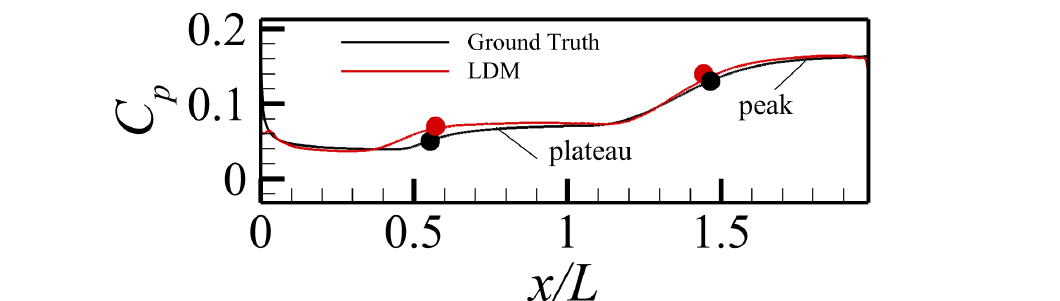}
	\caption{Comparison of wall pressure of the ground truth and LDM with $ Ma $ = 6.61,  $ Re   = 4.92\times 10^6$,  $ \alpha  $ = 11.1°, $ T_{\infty}$ = 113 K and  $ T_w /T_0 $ = 0.80. Closed circles: separation and reattachment points.}
	\label{figs:cp2}
\end{figure}

Figure~\ref{figs:cp2} compares the surface pressure coefficient $C_p$ between the ground truth and LDM at $ Ma $ = 6.61,  $ Re   = 4.92\times 10^6$,  $ \alpha  $ = 11.1°, $ T_{\infty}$ = 113 K and  $ T_w /T_0 $ = 0.80. It can be observed that the surface pressure begins to increase upstream of the separation point, which is governed by the free interaction process \cite{chapman1958investigation,hao2021occurrence}. Following the initial upstream rise, a plateau region forms. Subsequently, in accordance with oblique shock theory, the pressure rises again near the reattachment point, reaching its peak value. Overall, the LDM prediction aligns well with the ground truth, except for minor discrepancies in the vicinity of the separation region and the peak wall pressure.

To summarize, this section demonstrates the successful application of the LDM to full-resolution hypersonic flow fields, highlighting the potential of generative models for tackling large-scale flow prediction problems.

%
%
%
\section{Conclusion}
\label{sec1}
This study investigates the performance of generative models across thermal diffusion and incompressible to hypersonic flow regimes, covering thermal distribution in plates with holes, airfoil flow fields in aerospace applications, and hypersonic flow over compression ramps. Both DDPM and LDM are applied to achieve high accuracy predictions, with the latent space model further improving training efficiency without compromising prediction quality. The main findings are summarized as follows:

(i) The generative models of DDPM and LDM are employed for thermal/flow fields predictions, with applications from the temperature distributions on a plate with holes by thermal diffusion to the velocity and pressure fields around airfoils in incompressible regimes and the hypersonic flow fields of compression corners. 

(ii) Compared to DDPM, the latent space diffusion model retains high prediction accuracy—for instance, reducing the temperature distribution prediction error for the plate with a central circular hole from 0.013471 (DDPM) out of 1, to 0.013267 (LDM)—while significantly cutting the training time required at the original data scale, thereby substantially lowering overall training cost.

(iii) The LDM achieves a separation length prediction that deviates by only 4.28\% from the DNS value when evaluated on hypersonic flow over a compression ramp. This further demonstrates the efficiency of operating in latent space for large-scale flow data and confirms the feasibility of employing generative models in latent space to accurately predict complex flow fields.

In conclusion, this work validates DDPM and LDM ‐ based generative models across the thermal diffusion problem, incompressible and hypersonic flow regimes, demonstrating their generality under diverse flow conditions. The successful application of diffusion models in latent space also highlights the potential of generative approaches for more complex flow‑prediction tasks.

\section*{Acknowledgements}

This work is supported by the Hong Kong Research Grants Council (no.15217622 and no. 15203724).

\section*{Conflict of Interest}

The authors have no conflicts to disclose.

\section*{Data Availability Statement}

The data that support the findings of this study are available from the corresponding author upon reasonable request. 

\section*{Declaration of generative AI and AI-assisted technologies in the writing process}

During the preparation of this work the authors used DeepSeek in order to improve language. After using this tool, the authors reviewed and edited the content as needed and take full responsibility for the content of the publication.

\section*{Appendix}

%
%

\bibliographystyle{elsarticle-num}
\bibliography{viscous-damping}
%
%
\end{document}